\documentclass[final,5p,times,twocolumn]{elsarticle}
\usepackage{amsmath,amssymb,amsfonts}
\usepackage{amsthm}
\usepackage{float}
\usepackage{algorithm}
\usepackage{mathtools}
\usepackage{bbm}
\usepackage{algpseudocode}
\usepackage{graphicx}
\usepackage[noindentafter]{titlesec}
\usepackage{textcomp}
\usepackage{lineno,hyperref}
\usepackage{subfigure}
\usepackage{booktabs}
\usepackage{graphicx}
\usepackage{multicol}
\usepackage{multirow}
\usepackage{tabularx}
\usepackage{bbding}
\usepackage{paralist}
\usepackage{pifont}
\usepackage{wasysym}
\usepackage{chngcntr}
\usepackage{amssymb}
\usepackage{graphics}
\usepackage[symbol]{footmisc}

\usepackage{wrapfig}
\renewcommand{\arraystretch}{1.1}
\DeclareMathOperator*{\argmax}{arg\,max}

\journal{Journal of Systems and Software}

\begin{document}
\begin{frontmatter}

\title{HUNTER: AI based Holistic Resource Management for Sustainable Cloud Computing 
}

\author{Shreshth Tuli\fnref{one}}
\author{Sukhpal Singh Gill \fnref{two}}
\author{Minxian Xu\fnref{three}}
\author{Peter Garraghan\fnref{four}}
\author{Rami Bahsoon\fnref{five}}
\author{Schahram Dustdar\fnref{six}}
\author{Rizos Sakellariou\fnref{seven}}
\author{Omer Rana\fnref{eight}}
\author{Rajkumar Buyya\fnref{nine}}
\author{Giuliano Casale\fnref{one}}
\author{Nicholas R. Jennings\fnref{one,ten}}
\fntext[one]{Department of Computing, Imperial College London, UK}
\fntext[two]{School of Electronic Engineering and Computer Science, Queen Mary University of London, UK}
\fntext[three]{Shenzhen Institutes of Advanced Technology, Chinese Academy of Sciences, China}
\fntext[four]{School of Computing and Communications, Lancaster University, UK}
\fntext[five]{School of Computer Science, University of Birmingham, UK}
\fntext[six]{Distributed Systems Group, Vienna University of Technology, Austria}
\fntext[seven]{Department of Computer Science, University of Manchester, UK}
\fntext[eight]{School of Computer Science and Informatics, Cardiff University, UK}
\fntext[nine]{Cloud Computing and Distributed Systems (CLOUDS) Laboratory,
School of Computing and Information Systems, The University of Melbourne, Australia}
\fntext[ten]{Loughborough University, UK\\
E-mail addresses: s.tuli20@imperial.ac.uk (S. Tuli), s.s.gill@qmul.ac.uk (S.S. Gill), mx.xu@siat.ac.cn (M. Xu), p.garraghan@lancaster.ac.uk (P. Garraghan), r.bahsoon@cs.bham.ac.uk (R. Bahsoon), dustdar@dsg.tuwien.ac.at (S. Dustdar), rizos@manchester.ac.uk (R. Sakellariou), ranaof@cardiff.ac.uk (O. Rana), rbuyya@unimelb.edu.au (R. Buyya), g.casale@imperial.ac.uk (G. Casale) and n.r.jennings@lboro.ac.uk (N.R. Jennings).}

\begin{abstract}
The worldwide adoption of cloud data centers (CDCs) has given rise to the ubiquitous demand for hosting application services on the cloud. Further, contemporary data-intensive industries have seen a sharp upsurge in the resource requirements of modern applications. This has led to the provisioning of an increased number of cloud servers, giving rise to higher energy consumption and, consequently, sustainability concerns. Traditional heuristics and reinforcement learning based algorithms for energy-efficient cloud resource management address the scalability and adaptability related challenges to a limited extent. Existing work often fails to capture dependencies across thermal characteristics of hosts, resource consumption of tasks and the corresponding scheduling decisions. This leads to poor scalability and an increase in the compute resource requirements, particularly in environments with non-stationary resource demands. To address these limitations, we propose an artificial intelligence (AI) based holistic resource management technique for sustainable cloud computing called HUNTER. The proposed model formulates the goal of optimizing energy efficiency in data centers as a multi-objective scheduling problem, considering three important models: energy, thermal and cooling. HUNTER utilizes a Gated Graph Convolution Network as a surrogate model for approximating the Quality of Service (QoS) for a system state and generating optimal scheduling decisions.  Experiments on simulated and physical cloud environments using the  CloudSim toolkit and the COSCO framework show that HUNTER outperforms state-of-the-art baselines in terms of energy consumption, SLA violation, scheduling time, cost and temperature by up to 12, 35, 43, 54 and 3 percent respectively.  
\end{abstract}

\end{frontmatter}

\section{Introduction}
\label{sec:intro}


Cloud computing has proven to be a reliable, cost-effective and scalable computing service choice to host and deliver software solutions for diverse industrial applications~\cite{berl2010energy}. Many businesses have migrated to cloud data centers (CDCs) to take advantage of on-demand, elastic and scalable resource provisioning, saving companies on capital investments and maintenance of in-house infrastructure~\cite{shuja2016sustainable}. The plethora of deployment choices offered by most cloud providers allows users to customize resources according to their objectives. However, the rise of AI and Internet of Things (IoT) applications in Industry 4.0~\cite{chun2018study} has led to an increase in the overall requirements of cloud resources. 
In particular, cloud providers, such as Amazon, Microsoft and Google, have witnessed heavy reliance and increase in the number of cloud data centers to fulfill the increasing demands of users~\cite{wankhede2020comparative}. 

A large amount of energy is required to run these cloud data centers efficiently. Specifically, there is a need to manage the cloud resources effectively to lower the energy consumption and help reduce the cost and carbon footprints. This demand often comes with high energy consumption, a major part of which is attributed to the cooling costs~\cite{shuja2016sustainable}. The cooling infrastructure of a CDC can consume almost the same level of energy as the computing nodes themselves~\cite{pakbaznia2009minimizing}. Major public cloud providers need to invest in large scale cooling infrastructures, making it an expensive exercise~\cite{cruze}. Producing holistic energy-aware models for resource management, which consider both cooling and computational costs, has been acknowledged as an important open problem~\cite{cruze}. Specifically, the research gap presents a need for task scheduling in CDCs that considers the energy, thermal and cooling costs as optimization objectives~\cite{shuja2016sustainable}.  

\textbf{Challenges.} The problem of providing holistic resource management for sustainable cloud computing is fundamentally challenging because the relationship between energy consumption, the computational infrastructure and the cooling system is complex. Another challenge is the coordination of the scheduling decisions for different tasks that considers both computing power and cooling power in tandem. Another factor to consider in minimizing energy consumption is the reduction in resource intensive or thermal hotspots that can degrade the performance of the system~\cite{cruze}. Consequently, as non-stationary workloads are required to be serviced, the cooling systems and hence the power and temperature metrics of hosts change dynamically~\cite{chaudhry2015thermal}. Furthermore, tasks running in a datacenter may be inter-dependent. This is a common when jobs are allocated to a cloud environment with each job consisting of multiple independent tasks and service level agreements (SLAs) being defined for each job. The overall response time and SLA violations would then be defined at the job level instead of being measured for each task. Moreover, in hybrid public-private clouds, the host machines have different resource capacities in terms of their CPU, RAM, disk and network capabilities. These issues further complicate scheduling as now the representation of the system also needs to capture the inter task dependencies and host heterogeneity. 

\textbf{Existing solutions.} Over the past few years many resource management techniques have been proposed that target SLA compliance and the improvement of Quality of Service (QoS)~\cite{shuja2016sustainable}. Specific solutions that target sustainable computing aim at leveraging monitored metrics like energy consumption and temperature of host machines~\cite{shuja2016sustainable}. Only a few solutions also consider the energy and cost implications of cooling solutions~\cite{cruze}. Most prior work presents meta-heuristic algorithms~\cite{cruze} and deep learning techniques~\cite{chaudhry2015thermal}. Most state-of-the-art models use meta-heuristic approaches like genetic algorithms or integer linear programming~\cite{cruze, mitec, hdic, ann, fareghzadeh2019toward}. Other recent methods use reinforcement learning (RL); specifically, the traditional tabular models like Q-Learning~\cite{padqn, sdaemmq}. However, such meta-heuristic and RL techniques require several exploratory decisions before updating their models, making them harder to adapt quickly in highly volatile settings, considering inter-task dependencies, thermal characteristics or converging quickly to a scheduling decision~\cite{tuli2021cosco}. All such features are crucial for a holistic solution for sustainable scheduling~\cite{shuja2016sustainable}.

\textbf{Background and new insights.}
As is common in most prior work, modelling the optimization variable, \textit{i.e.} the scheduling decision, as a placement matrix does not capture the inter-task dependencies well~\cite{zhao2021distributed}. An improved approach is to use geometric modelling of the scheduling decisions, particularly as a graph, as it enables structure specific extraction of the system state information. Recently proposed Artificial Intelligence (AI) techniques, such as graph neural networks or graph encoders, can efficiently capture such geometric data~\cite{ruiz2020gated, wu2020comprehensive}. One such network, called Gated Graph Convolution Network (GGCN) enables aggregation of the graph node information using convolution operations and message passing~\cite{ruiz2020gated}, making it suitable to model distributed computing network as a graph. This enables a more versatile optimization approach that also takes into account the task hierarchy and edge-cloud hierarchy, not considered in most prior work~\cite{tuli2021cosco, hdic}. We use a GGCN model as a surrogate of the QoS objective scores allowing us to swiftly run placement optimization. Such a surrogate model enables us to quickly get the QoS score for an input (scheduling decision) without actually executing it in the physical environment, saving us time and cost. Such deep surrogate models are commonly used in the literature~\cite{ann, sdaemmq, tuli2021cosco}.

\textbf{Our contributions.} In this work, we significantly extend our previous work~\cite{cruze} by proposing a \textbf{\textit{H}}olistic reso\textbf{\textit{U}}rce ma\textbf{\textit{N}}agemen\textbf{\textit{T}} technique for \textbf{\textit{E}}nergy-efficient cloud computing using a\textbf{\textit{R}}tificial intelligence, called \textbf{\textit{HUNTER}}. The proposed method uses a GGCN network as a QoS surrogate to optimize the scheduling decision for a hybrid public-private cloud environment. It uses \textit{performance to power ratio} as a heuristic to explore the scheduling search space that enables us to significantly reduce scheduling time. In our previous work~\cite{cruze}, we proposed a holistic management technique for cloud resources and established a relationship between replication and service consolidation to improve the energy-efficiency and cut the carbon footprint.  However, our previous work did not deal with heterogeneous resources and dynamic workloads. In this work, we extend existing thermal and energy consumption models to also include the cooling overheads~\cite{chaudhry2015thermal}. 
Further, to prevent excessive scheduling overheads, we use performance to power ratio as a heuristic to significantly reduce the time to converge to a scheduling decision~\cite{zhang2013towards}. To adapt in volatile scenarios, we periodically adjust the weights of the deep surrogate model using backpropagation.

The contributions of this research work are summarized as:
\begin{compactitem}  
\item We propose a novel energy-efficient resource management approach (HUNTER) that uses \textit{GGCN as a deep surrogate model} for quick QoS estimation and three sustainability models, \textit{viz}, thermal, energy and cooling.
\item Extensive experiments on simulated (using the CloudSim~\cite{calheiros2011cloudsim} toolkit) and physical cloud testbeds (using the COSCO~\cite{tuli2021cosco} framework) show that the proposed model outperforms state-of-the-art schedulers for sustainable computing.
\item HUNTER gives the \textit{best} energy consumption, SLA violation, cost and temperature by 11.90\%, 35.41\%, 53.86\% and 3.47\% respectively. HUNTER achieves this with 42.78\% lower scheduling overheads compared to the best baseline.
\end{compactitem}

The rest of the article is organized as follows. Section~\ref{sec:related} overviews the related work. Section~\ref{sec:hunter} presents the HUNTER scheduler. Performance evaluation is carried out in Section~\ref{sec:experiments} with additional results and analysis in Section~\ref{sec:analyses}. Section~\ref{sec:conclusion} concludes the paper and proposes future work. 

\begin{table*}[]
\centering
\renewcommand{\arraystretch}{1.1}
\caption{Comparison of HUNTER with related work (\checkmark means that the corresponding feature is present).}
\label{tab:related_work}
\centering
\resizebox{\textwidth}{!}{
\begin{tabular}{@{}lccccccccccc@{}}
\toprule 
\multirow{2}{*}{Work} & \multirow{2}{*}{Holistic} & \multirow{2}{*}{Dynamic} & Technical & \multicolumn{3}{c}{Sustainability Models} & \multicolumn{5}{c}{QoS and other Optimization Parameters}\tabularnewline
 &  &  & Approach & Energy & Thermal & Cooling & Temperature & Time & SLA Violation Rate & Cost & Energy\tabularnewline
\midrule 
TOPSIS~\cite{arianyan2015novel} & \checkmark &  & Threshold Based & \checkmark & \checkmark &  & \checkmark &  &  &  & \checkmark\tabularnewline
MALE~\cite{liang2020memory} & \checkmark &  & Memory Mapping & \checkmark & \checkmark &  & \checkmark &  &  &  & \checkmark\tabularnewline
CRUZE~\cite{cruze} & \checkmark &  & Cuckoo Optimization & \checkmark & \checkmark & \checkmark & \checkmark & \checkmark & \checkmark & \checkmark & \checkmark\tabularnewline
MITEC~\cite{mitec} &  & \checkmark & Genetic Algorithm & \checkmark & \checkmark &  & \checkmark &  &  &  & \checkmark\tabularnewline
PADQN~\cite{padqn} &  & \checkmark & Deep Q Learning & \checkmark &  &  & \checkmark & \checkmark &  &  & \checkmark\tabularnewline
ANN~\cite{ann} &  & \checkmark & Neural Network & \checkmark &  &  &  & \checkmark &  &  & \checkmark\tabularnewline
SDAE-MMQ~\cite{sdaemmq} &  & \checkmark & Autoencoders & \checkmark &  &  &  & \checkmark & \checkmark &  & \checkmark\tabularnewline
HDIC~\cite{hdic} & \checkmark & \checkmark & NARX Network & \checkmark & \checkmark &  & \checkmark &  &  &  & \checkmark\tabularnewline
\textbf{HUNTER} & \checkmark & \checkmark & Surrogate Modelling & \checkmark & \checkmark & \checkmark & \checkmark & \checkmark & \checkmark & \checkmark & \checkmark\tabularnewline
\bottomrule 
\end{tabular}}
\end{table*}

\section{Related Work}
\label{sec:related}

A significant amount of research has been devoted to the area of resource management in cloud computing. Table~\ref{tab:related_work} summarizes the comparison of HUNTER with existing works based on important key features. Given our scope of holistic management of resources particularly focusing on sustainability, we classify the state-of-the-art work into two main categories: 1) meta-heuristic methods (rows 1-3) and 2) reinforcement learning models (rows 4-8). The `holistic' column represents whether the approach provides  an end-to-end solution for scheduling, considering all parameters for sustainable cloud computing~\cite{cruze}. The `dynamic' column represents whether the technique adapts on-the-fly for non-stationary environments.

\textbf{Meta-Heuristic Methods.} Our previous work, CRUZE~\cite{cruze}, aimed to reduce the total cloud energy consumption whilst maximizing reliability of the system. It utilizes efficient design models with respect to energy, reliability, capacity and cooling. To generate a scheduling decision, CRUZE uses a Cuckoo optimization approach. Another work, FECBench~\cite{barve2019fecbench} provides performance interference prediction models for services of cloud providers to develop resource management techniques. The authors have constructed a process pipeline to construct multi-resource stressors using machine learning. To minimize prohibitive profiling costs, the authors have explored multi-dimensional resource metrics with minimal experimental runs using design of experiments (DoE) that significantly minimizes prohibitive profile cost. In similar efforts, the MALE algorithm~\cite{liang2020memory} was recently introduced to minimize energy consumption in a cloud datacenter by reducing memory consumption and contention. This is achieved by mapping memory requirements of virtual machines to cloud hosts using a predefined best-fit criteria. Similarly, TOPSIS~\cite{arianyan2015novel} presents a set of heuristics to significantly reduce energy consumption using thermal features that are recorded from cooling devices and servers. This approach uses a threshold based load-balancing technique to prevent thermal hotspots and minimize failures due to overheating. A similar work, MITEC~\cite{mitec} uses a genetic algorithm to optimize scheduling decisions and updates the energy and thermal models to tune the fitness scores for each allocation decision. Other works in this category propose autonomic cloud resource management mechanisms for the execution of batch and interactive workloads by leveraging the multiple resource layers and host heterogeneity to reduce energy consumption~\cite{shuvo2021energy, butt2019cloud, kumar2019energy}. However, most methods in this category including CRUZE, Ella-W and GRANITE do not adapt in volatile settings. Still, we include the CRUZE and MITEC methods as a baseline in our experiments to represent this category.

\textbf{Reinforcement Learning.} In recent years, several machine learning (ML) based schedulers have been proposed that aim to optimize energy consumption of CDCs. Reinforcement learning (RL) is an sub-field within ML that models the system as an interactive environment using QoS parameters to dynamically modify the scheduling policy~\cite{sutton2018reinforcement}. One of the most versatile RL techniques is the deep Q learning (DQL). Here, a deep neural network is used to estimate a long-term reward (commonly referred to as the Q value) for each state. Many recent works, for instance PADQN~\cite{padqn} and SDAE-MMQ~\cite{sdaemmq}, formulate the scheduling problem as a RL problem and utilize deep Q learning to produce task placement decisions~\cite{zeng2019resource}. Here, the decision is modelled as the state of the RL system with actions as task migration or allocation decisions. Each action changes the model state and gives a reward in the form of a QoS score. More advanced DQL based approaches use sophisticated neural networks to predict the Q values for each scheduling decision. SDAE-MMQ uses a stacked denoising autoencoder as a value network and MiniMax-Q instead of vanilla Q-learning~\cite{sdaemmq}. Another work, HDIC~\cite{hdic}, uses a nonlinear auto-regressive network with exogenous inputs (NARX) as a value network. Advanced neural models typically take a long time to train and infer Q values for large-scale state inputs. Other works directly use a deep neural network to produce a task allocation or migration decision~\cite{witanto2018adaptive}. For instance, a recent ANN approach uses an artificial neural network to produce a softmax output for each task~\cite{ann}. Taking the argmax for each output gives us the scheduling decision for each task.  The ANN is trained using a supervised learning framework with actions being rewarded using QoS metrics like energy consumption and execution time. Such approaches usually scale well with the number of tasks or hosts in the system, but are unable to capture inter-task dependencies to efficiently handle task placement. Moreover, Q-learning based methods are known to be slow to adapt in volatile settings~\cite{tuli2021cosco}. We use the PADQN, ANN, SDAE-MMQ and HDIC methods as baselines in our experiments as these are empirically the best methods this category.

These approaches focus on particular perspectives of cloud resources management, e.g. computing or network. Unlike these works, our approach considers resource management in a holistic manner by considering energy, thermal and cooling characteristics whilst reducing scheduling and task migration overheads.

\begin{figure*}[t]
	\centering
	\includegraphics[width=\linewidth]{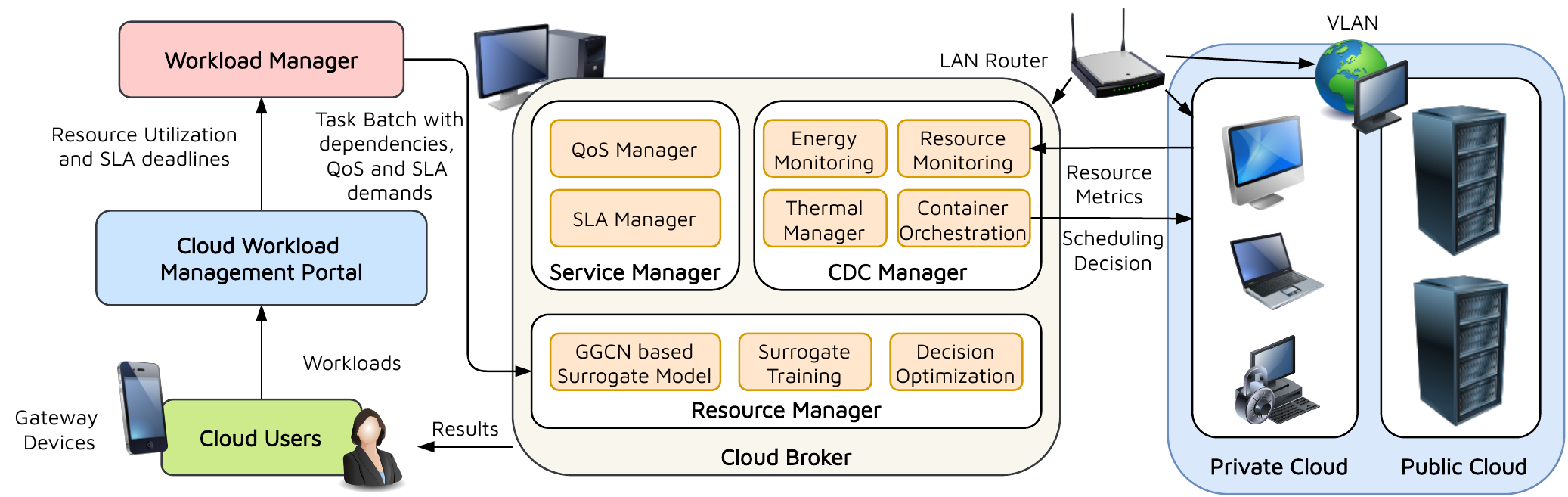}
	\caption{System Model of HUNTER}
	\label{fig:system}
\end{figure*}

\section{The HUNTER Scheduler}
\label{sec:hunter}

\subsection{System Model}
Figure~\ref{fig:system} shows the system model considered in this work. Motivated from prior work~\cite{madej2020priority, tuli2021cosco}, we consider the following components of the CDC.
\begin{compactitem}  
\item \textbf{Cloud Users:} The users share workloads as jobs to the CDC (more details in Section~\ref{sec:workload_model}). The data is collected using IoT sensors and passed on to the CDC using gateway devices like smartphones and tablets~\cite{tuli2021cosco}.
\item \textbf{Cloud Workload Management Portal:} A graphical user interface for cloud users to interact with the system for the submission of workloads along with their SLA and QoS details.  
\item \textbf{Workload Manager:} Initially, the workload manager processes all the incoming workloads. An admission controller realizes all workloads as container instances~\cite{cruze, tuli2021cosco}. 
\item \textbf{Cloud Broker:} The central cloud server that allocates incoming jobs to various compute resources (cloud worker nodes). It consists of the following components:
\begin{compactitem}
\item \textit{Service Manager:} Contains two elements, SLA and QoS managers that manage the heterogeneous cloud services while processing workloads. The QoS manager contains the information about QoS requirements for different workloads, while the SLA manager contains the information about an agreement signed between a cloud user and a provider based on QoS requirements. 
\item \textit{CDC Manager}: Continually monitors the resource utilization of all active tasks and hosts in the system.  It also monitors the QoS parameters (including the energy and thermal characteristics of cloud hosts) and also performs the task allocation and migration. In this work, we assume tasks as container instances and task migration as the transfer and restoration of container checkpoints.
\item \textit{Resource Manager:} Decides the schedule for each task in the system. The resource manager includes the sustainability models for energy, thermal and  cooling parts of the CDC. For resource scheduling, the manager contains a GGCN based surrogate model that estimates QoS parameters. It performs training and on-the-fly tuning of the GGCN model to adapt in non-stationary settings. This manager also runs an exploration strategy that checks the QoS scores for a set of allocations and chooses the best one as the scheduling decision (more details in Section~\ref{sec:scheduling}). 
\end{compactitem} 
\item \textbf{Cloud Hosts:}  The cloud broker is connected to a heterogeneous set of worker nodes. Some nodes are present in the same Local Area Network (LAN) as the broker, called the private cloud. Others are present in a geographically distant location and connected using a virtual LAN (VLAN). As is common practice, we assume that private-cloud nodes are resource constrained but offer low latency services, and public-cloud nodes have abundant resources but have high communication latency.
\end{compactitem}
The HUNTER scheduler resides as the Resource manager in the Cloud Broker, taking tasks as inputs from the Workload Manager (see Figure~\ref{fig:system}). HUNTER uses resource metrics from the Resource Monitor and executes scheduling decision through Container Orchestration (as tasks are realized as containers in our system). 

\subsection{Workload Model and Problem Formulation}
\label{sec:workload_model}

As common in prior work, we assume that generating scheduling decisions is a discrete-time control problem~\cite{tuli2021cosco, basu2019learn}. We divide the timeline into equal duration intervals, with the $t$-th interval denoted as $I_t$ (starting from $i=0$). We assume a fixed number of host machines and denote the set of cloud hosts by $H$. The workloads are in the form of jobs $J_t$, where each job $j_i \in J_t$ is composed of multiple tasks $j_i = \{t_0, \ldots, t_{|j_i|}\}$. There are no precedence constraints among tasks that belong to the same job, but the QoS metrics are calculated at the job level instead of the task level. Thus, it is important to consider inter-task dependencies while scheduling.  All new jobs created at the interval $I_t$ are denoted as $N_t$, with all active jobs being denoted as $A_t$. A job is considered to be active if at least one task of that job is being executed in the cloud environment. If no task of a job $j \in N_t$ can be allocated to a cloud node then it is added to a wait queue $W_t$. All created jobs that are not active and are not in the wait queue are considered to be completed and we can calculate their metrics like response time and SLA violation.

We consider the problem of maximizing the QoS objective score accumulated across all intervals in a bounded time experiment. We denote the QoS score for interval $I_t$ by $O_t$ and consider a total $n$ intervals in an experiment. We denote the utilization metrics of all hosts in interval $I_{t-1}$ as $U_t$. Now using $U_t$, we need to predict a scheduling decision $S_t$. All tasks for jobs in $N_t \cup W_t \cup A_t$ are called feasible tasks. Thus, the problem can be formulated as:
\begin{equation}
\label{eq:problem}
\begin{aligned}
& \underset{S_t}{\text{maximize}}
& & \sum_{t=0}^{n} O_t \\
& \text{subject to}
& & \forall\ t, S_t : P_t \cup Q_t \rightarrow H, \\
&&& \forall\ t, P_t = \text{ set of feasible tasks in }N_t \cup W_t \cup A_t, \\
&&& \forall\ t, Q_t = \text{ set of active tasks in the system.}
\end{aligned}
\end{equation}

In the rest of the discussion we consider these symbols only for a single interval and drop the $t$ subscript for notational convenience.

\subsection{Sustainability Models}
In this work, to decouple the different aspects of sustainable resource management~\cite{cruze}, we have designed three different models: energy, thermal and cooling. For completeness, we reproduce the formulae from our prior work~\cite{cruze}, with necessary adaptations for the new formulation.

\subsubsection{Energy Model} 
This model is designed to encapsulate all parameters related to the energy consumption, ranging from compute devices to the cooling components~\cite{mirhoseininejad2020joint}. The total energy of a CDC is calculated as
\begin{equation}
\label{eq:energy}
E\textsubscript{Total}=E\textsubscript{Computing}+E\textsubscript{Cooling}.
\end{equation}

The computing system consists of hosts and its energy consumption includes that of the different components like CPU, RAM, disk, network and peripherals. Thus, $E\textsubscript{Computing}$ can be defined as
\begin{equation}
E\textsubscript{Computing}=E\textsubscript{Processor}+E\textsubscript{Storage}+E\textsubscript{Memory}+E\textsubscript{Network}+E\textsubscript{Extra}.
\end{equation}

\textbf{Processor.} Here, $E\textsubscript{Processor}$ represents the processor’s energy consumption, which is calculated by adding the idle and dynamic consumption of all cores. Thus, 
\begin{equation}
E\textsubscript{Processor}=\sum_{r=1}^{cores} E_\text{dynamic}^{r} + E_\text{idle}^{r},
\end{equation}
where $E_\text{dynamic}^{r}$ and $E_\text{idle}^{r}$ are the dynamic and idle energy consumption of the $r$-th core. Here, $E\textsubscript{dynamic}$ is calculated using 
\begin{equation} 
E\textsubscript{dynamic}=\frac{E_\text{dynamic}^\text{linear}+E_\text{dynamic}^\text{non-linear}}{2}.
\end{equation}
$E_\text{dynamic}^\text{linear}$ is calculated as
\begin{equation}
E_\text{dynamic}^\text{linear}=CV^2 f,
\end{equation}
where $C$ is CPU capacitance, $f$ is CPU clock frequency, and $V$ is CPU voltage. $E_\text{dynamic}^\text{non-linear}$ is calculated using 
\begin{equation}
E_\text{dynamic}^\text{non-linear}(h_j) = \mu_1 \cdot U_j + \mu_2 \cdot U_j^2,
\end{equation}
where $\mu_1$ and $\mu_2$ are non-linear model parameters and $U_j$ is CPU utilization of host $h_j$.

\textbf{Storage.} $E\textsubscript{Storage}$ represents the energy consumption of storage devices to store data. The data read and write operations account for the energy consumption in such devices
\begin{equation}
E\textsubscript{Storage}=E\textsubscript{ReadOperation}+E\textsubscript{WriteOperation}+E\textsubscript{Idle}.
\end{equation}
$E\textsubscript{Memory}$ represents the energy consumption of the main memory (RAM/DRAM) and cache memory (SRAM), which is calculated using
\begin{equation}
E\textsubscript{Memory}=E\textsubscript{SRAM}+E\textsubscript{DRAM}.
\end{equation}

\textbf{Network.} $E\textsubscript{Network}$ represents the energy consumption of networking equipment such as routers, switches and gateways, LAN cards, etc., and is calculated as
\begin{equation}
E\textsubscript{Network}=E\textsubscript{Router}+E\textsubscript{Switches}+E\textsubscript{Gateways}+E\textsubscript{LAN cards}.
\end{equation}

\textbf{Peripherals.} $E\textsubscript{Extra}$ represents the energy consumption of other parts, including the current conversion loss and others and is calculated as
\begin{equation}
E_\text{Extra}=E_\text{motherboard} + \sum_{f\in F} E_\text{connector}^{f}
\end{equation}
where $E\textsubscript{motherboard}$ is energy consumed by motherboard(s) and $\sum_{f\in F}
E_\text{connector}^{f}$ is energy consumed by a connector (port) running at the frequency \textit{f}, where the set of port frequencies is denoted by $F$.
\subsubsection{Cooling Model} 
In the cooling model, $E\textsubscript{Cooling}$ denotes the energy consumed by cooling devices to maintain the temperature of a cloud datacenter, which is calculated using
\begin{equation}
E\textsubscript{Cooling}=E\textsubscript{AC}+E\textsubscript{Compressor}+E\textsubscript{Fan}+E\textsubscript{Pump},
\end{equation}
where $E\textsubscript{AC}$ is the energy consumption of the air-conditioner inside the cloud-datacenter, $E\textsubscript{Compressor}$ is the energy used by the compressor, $E\textsubscript{Fan}$ is that of the fans attached to the radiators and $E\textsubscript{Pump}$ is that of the pump within the all-in-one (AIO) water cooling solution.

\subsubsection{Thermal Model } 
To design the thermal model, we use the Computer Room Air Conditioning (CRAC) model and RC (where R and C are thermal resistance (k/w) and heat capacity (j/k) of the host respectively) used as a time-constant to estimate the temperature of the CPU  for each host ($T\textsubscript{cu}$)~\cite{cruze, chaudhry2015thermal}. Thus,
\begin{equation}
\label{eq:temp}
T\textsubscript{cu}=PR+Temp\textsubscript{inlet} + T\textsubscript{initial} * e\textsuperscript{-RC},
\end{equation}
where the inlet temperature ($Temp\textsubscript{inlet}$) is calculated using CRAC model ($T\textsubscript{cu}$); the RC model is used to calculate CPU temperature ($T\textsubscript{CPU}$); $P$ is the dynamic power of host. $T\textsubscript{inital}$ is the initial temperature of the CPU, which is taken as the ambient temperature of the datacenter~\cite{chaudhry2015thermal}. 

The detailed description of the thermal model and the various metrics is given in our previous works~\cite{cruze}.

\subsection{GGCN based Surrogate Model}
\label{sec:surrogate}
\begin{figure}[t]
    \centering
    \includegraphics[width=\linewidth]{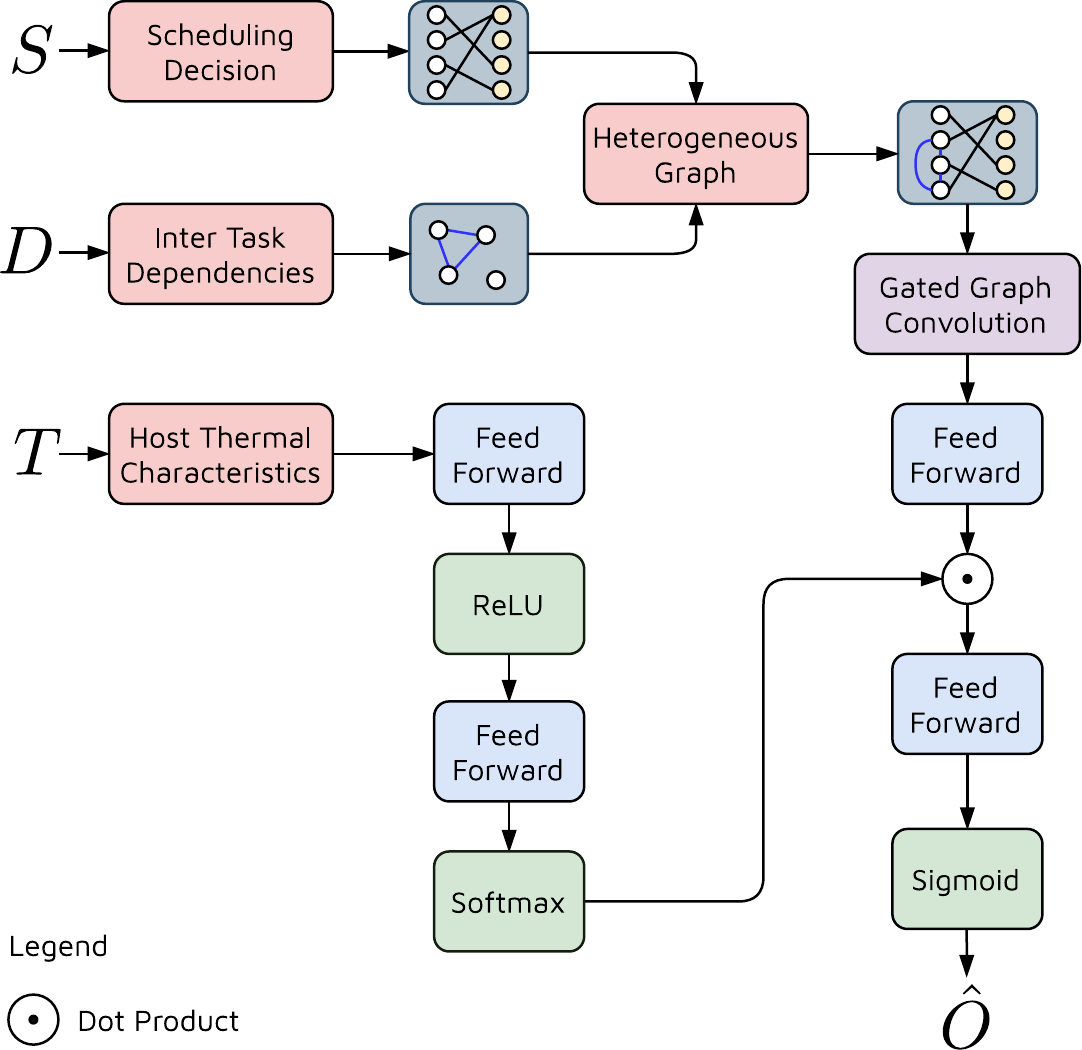}
    \caption{GGCN based surrogate model of HUNTER. The three inputs to the model and the graph structure are shown in red. Feed-forward and graph convolution operations are shown in blue and purple respectively. All activations are shown in green and all data structures are shown in grey.}
    \label{fig:model}
\end{figure}


As described in Section~\ref{sec:intro}, we model the inputs of our surrogate model using a geometric graph representation (see Figure~\ref{fig:model}). We form two graphs $D$ and $S$. The former represents the inter task dependencies and the latter represents the bi-partite graph corresponding to the scheduling decision. $D = (V_D, E_D)$, where $V_D$ denotes the tasks as nodes and $E_D$ denotes the inter-task dependency in terms of the jobs they belong to as undirected edges. Each task has a feature vector corresponding to the Instructions per Second (IPS), RAM, Disk and Bandwidth consumption. The RAM, Disk and Bandwidth consumption also include the read and write speeds of such tasks. Each edge $(t_p,t_q)$ in $E_D$ is an unordered pair such that tasks $t_p$ and $t_q$ belong to the same job $j \in J$. $S = (V_S, E_S)$ is a bi-partite graph with nodes of two types: tasks and hosts. The edges of the graph $(t_p, h_r) \in E_S$ correspond to the allocation decision of the current state, where task $t_p$ is allocated to host $h_r$. Similar to $D$, each task has a feature vector corresponding to the IPS, RAM, Disk and Bandwidth consumption. The feature vectors of the host machines consist of the IPS, RAM, Disk and Bandwidth consumption and capacities. 

To perform graph convolutions, we combine $S$ and $D$ into a single heterogeneous graph, where the edge set now becomes $E_D \cup E_S$, such that each edge now also has an edge type (task dependence or allocation relation). This hetero-graph is then sent to the GGCN model to run convolution operations. The GGCN model executes convolutions across the edges of the graph where the convolution operations are weighted using a Gated Recurrent Unit (GRU). The convolution operations allow the model to share information across different tasks and hosts whilst inferring a latent representation of the scheduling decision. This information sharing helps the downstream operations to explicitly leverage the inter-task dependencies and the implications of an allocation or migration decision on host utilization characteristics. The GRU based weighting allows the model to be flexible with respect to the extent to which the feature vectors of hosts and tasks should be combined. Formally, the message passing leads to \textit{graph-to-graph} updates
\begin{align}
\begin{split}
    r_i^{0} &= \mathrm{Tanh} \left( W\ e_{i} + b \right),\\
    x^k_i &= \sum_{j \in n(i)} W^k r_{j}^{k-1} ,\\
    r^k_{i} &= \mathrm{GRU} \left( r^{k-1}_i, x^{k}_{i} \right).\\
\end{split}
\end{align}
Here, $W, b$ are parameters of the feed-forward layer within the GGCN network, $e_i$ is the feature-vector (described previously) of a node $i \in V_D \cup V_S$ in the heterogeneous graph and $k$ varies from 1 to $p$. Also, the messages for task $i$ are aggregated over one-step connected neighbors $n(i)$ over the $p$ convolutions, resulting in an embedding $r^{p}_{i}$ for each task node in the graph. The stacked representation for all tasks is represented as $r^p$. Convolutions across these edge types help as the dependence of task utilization characteristics with the allocated hosts is the maximum and much lower for hosts on which the task is not allocated. This allows the scheduling decision to properly manage task and hosts utilization characteristics while also considering the changing demands of tasks. We generate the graph encoding $e^S$ by passing $r^p$ through a feed-forward layer as
\begin{equation}
    e^S = \mathrm{FeedForward}(r^p).
\end{equation}

\begin{figure}[t]
    \centering
    \includegraphics[width=0.7\linewidth]{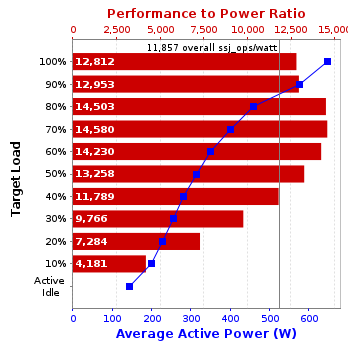}
    \caption{A sample power to load curve for a cloud server. Reproduced with permission. Source: SPEC benchmark power profile repository \url{https://www.spec.org/power_ssj2008/results/res2021q2/power_ssj2008-20210528-01098.html}.}
    \label{fig:sample}
\end{figure}

We also capture the thermal-characteristics of the host machines in terms of their current temperatures ($T_\text{cu}$) and the power to load profile\footnote{We use Standard Performance Evaluation Corporation (SPEC) power consumption models to generate power to load performance curves of our cloud hosts. URL: \url{https://www.spec.org/power_ssj2008/}} (see Figure~\ref{fig:sample} for a sample performance to power profile). We model the thermal profile and current temperature of all hosts as vectors ($Temp$) and pass through a FeedForward network
\begin{equation}
    e^T = \mathrm{ReLU}(\mathrm{FeedForward}(\mathrm{ReLU}(\mathrm{FeedForward}(Temp)))).
\end{equation}

We then use Bahdanau style self-attention~\cite{bahdanau2015neural} to generate an estimate of the QoS objective. This allows the model to focus only on those hosts that can potentially become thermal hotspots. \begin{equation}
    \hat{O} = \mathrm{Sigmoid}(\mathrm{FeedForward}(e^S \cdot \mathrm{Softmax}(e^T))).
\end{equation}
The sigmoid operator allows us to generate an output within $[0, 1]$ to enable training with normalized QoS scores. The GGCN model is agnostic to the QoS objective in general; however, in our experiments we use energy, temperature and SLA violation rates to train and fine-tune the model (see Section~\ref{sec:scheduling}). To train the GGCN model, we use the Mean-Square-Error (MSE) loss between the predicted and ground-truth QoS scores.

\begin{figure}[]
	\centering \setlength{\belowcaptionskip}{-10pt}
    \includegraphics[width=\linewidth]{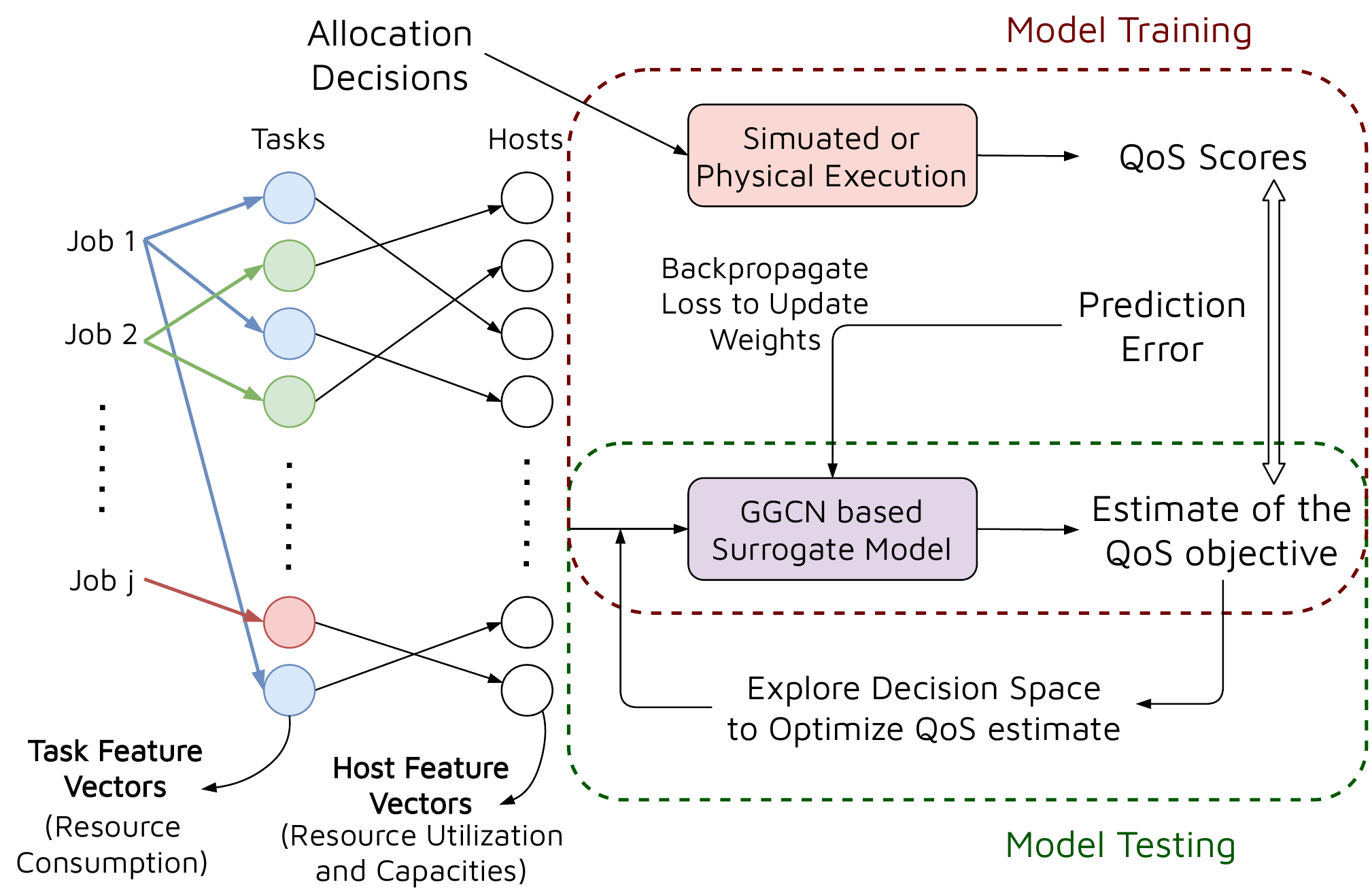}
	\caption{Graphical representation of the GGCN model. }
	\label{fig:overview}
\end{figure}

\subsection{Using GGCN Model for Scheduling}
\label{sec:scheduling}
We now present the \textit{modus operandi} of the proposed HUNTER resource scheduling technique using the GGCN network as a surrogate model (summarized in Algorithm~\ref{alg:ggcn}). Figure~\ref{fig:overview} shows a graphical representation of the HUNTER scheduler.  The input for the scheduler is the resource utilization metrics of the hosts and tasks that need to be allocated or migrated and the thermal characteristics of the cloud hosts. These are obtained from the resource monitoring and the thermal management services in the cloud broker (see Figure~\ref{fig:system}). These metrics are then combined to generate the $S$ and $D$ graphs and the $T$ vector as described in Section~\ref{sec:surrogate}. The allocation in the $S$ graph is obtained from the scheduling decision of the previous interval for $I_t$ for $t>0$ and a random allocation otherwise (line~\ref{line:monitor} in Alg.~\ref{alg:ggcn}). Thus, from the input $[S, D, T]$ we obtain an estimate of the QoS score $\hat{O}$. Now to generate a scheduling decision, we choose subsets of all tasks and hosts, each of size $K$. We sort the tasks based on the power consumption of their hosts, breaking ties using the CPU utilization and consider top $K$ such tasks (line~\ref{line:tasks} in Alg.~\ref{alg:ggcn}). Now, for each such high energy-resource consumption task, we choose a target host where it can be migrated. To do this, we sort all hosts in terms of the \textit{performance to power ratio} and choose the bottom $K$ hosts (line~\ref{line:hosts} in Alg.~\ref{alg:ggcn}). Now, we consider all $K\times K$ combinations and choose the host $h$ that maximizes the objective score estimate obtained from the surrogate model. We denote the updated $S$ graph with task-host allocation $(t, h)$ by $S(t, h)$ (line~\ref{line:check} in Alg.~\ref{alg:ggcn}).

To choose the value of the $K$ parameter, we leverage the network transfer constraints. Consider the router bandwidth (in MB/s) in a CDC to be denoted by $B$ and the size of a scheduling interval (in seconds) by $I_S$. Also, let us denote the minimum size of a container in the CDC in MB by $S_C$. Then we define
\begin{equation}
    K = \frac{B \times I_S}{S_C}.
\end{equation}
Intuitively, this denotes the upper-bound of the number of tasks that can be migrated in the scheduling interval. If we try to migrate more than $K$ tasks in any decision, the tasks would take more than $I_S$ seconds, rendering the migration useless as new decisions are taken every $I_S$ seconds. For a typical setting with bandwidth of 100-MB/s, scheduling interval of 300-seconds and container size of 3000-MB, $K = 10$ allowing us to check only 100 task-to-host allocations. This can be significantly more efficient than checking all task-host combinations, which may be in thousands. 

\begin{algorithm}[t]
    \begin{algorithmic}[1]
    \Require
    \Statex Pre-trained GGCN model $f$
    \Statex Convergence threshold $\epsilon$
    \Statex Consideration parameter $K$; Learning rate $\gamma$
    \Procedure{HUNTER}{s}
    \State $S, D, T \gets ResourceMonitor()$ \label{line:monitor}
    \State $\hat{O} \gets f([S, D, T])$
    \State $Tasks \gets$ get top $K$ tasks based on power consumption \label{line:tasks}
    \State $Hosts \gets$ get bottom $K$ hosts based on performance to power ratio \label{line:hosts}
    \State \textbf{for} ($t \in Tasks$) \textbf{do}
    \State \hspace{\algorithmicindent} host $\gets \argmax_{h \in Hosts} f([S(t, h), D, T]) $\label{line:check} 
    \State \hspace{\algorithmicindent} \textbf{if} (allocation of $t$ to $h$ is feasible) \label{line:if}
    \State \hspace{\algorithmicindent} \hspace{\algorithmicindent} Allocate or migrate $t$ to $h$.
    \State \hspace{\algorithmicindent} \textbf{else}
    \State \hspace{\algorithmicindent} \hspace{\algorithmicindent} Add $t$ to wait queue.
    \State \hspace{\algorithmicindent} \textbf{end if}
    \State \textbf{end for}
    \State $O = 1 - (\alpha \cdot AEC + \beta \cdot AT + \gamma \cdot SLAV)$ \label{line:objective}
    \State Datapoint $\gets ([S, D, T], O)$
    \State Backpropagate $f$ using Datapoint and MSE loss \label{line:backprop}
    \EndProcedure
    \end{algorithmic}
\caption{HUNTER Scheduler}
\label{alg:ggcn}
\end{algorithm}

The motivation behind using the performance-to-power ratio to sort hosts is as follows. Consider the sample performance to power ratio profile shown in Figure~\ref{fig:sample}. This ratio indicates the amount of CPU computational performance we get for each watt of power consumed. It is apparent that this ratio is highest for 70\% CPU load and reduces for higher or lower CPU loads. Most cloud servers have similar trends in their power profiles, with a sweet spot around 70-80\%. Higher CPU load can be easily avoided by  capping the constraint checker in the scheduler to not allocate tasks to host with 80\% CPU load. However, choosing the optimal target host for lower CPU loads is challenging due to the heterogeneity of the host power profiles.  This key insight of using the performance-to-power ratio allows us to minimize the number of migrations as well as the time to explore different scheduling options. 

Further, to adapt in volatile settings, at every scheduling interval we train the neural approximator using back-propagation. To do this, we obtain the latest QoS objective score from the QoS manager of the cloud broker and fine-tune the weights of the GGCN model by back-propagating the MSE loss between the predicted and true QoS scores. The ground-truth QoS score is obtained as (lines~\ref{line:objective}-\ref{line:backprop} in Alg.~\ref{alg:ggcn})
\begin{equation}
    O = 1 - (\alpha \cdot AEC + \beta \cdot AT + \gamma \cdot SLAV),
\end{equation}
where $AEC$, $AT$ and $SLAV$ denote the average normalized energy consumption, average normalized temperature and SLA violation for the leaving tasks in the previous interval. Here, $\alpha, \beta, \gamma$ are convex-combination weights. To minimize the metrics of energy, temperature and SLA violations, we maximize $O$. Continuous training of the model allows it to quickly adapt to dynamic workloads and also consider changing scheduling decisions and use these as well to consider task migrations and allocations. As shown in line~\ref{line:if} in  Alg.~\ref{alg:ggcn}, for each container, we check if allocation to the host corresponding to the maximum QoS score of the GGCN model is feasible, if yes we allocate/migrate the container to this host else add it to the wait queue to be processed in the next interval.

\textbf{Computational Complexity.} Assuming that the inference of a deep neural network is an $O(1)$ operation, we provide the computational complexity in the Big-O notation. Assume $p = |P\cup Q|$ active tasks and $q = |H|$ hosts in the system. Selecting $K \leq q$ hosts and tasks from these sets based on the previously described metric is an $O(q \log{K})$ and $O(p \log{K})$ operation. Checking all $K\times K$ task to host allocations is a $O(K^2)$ operation. Overall, the computational complexity of checking the various allocation choices is $O(K^2 + q \log{K} + p \log{K}) = O(p + q)$ as $K$ is a fixed constant (hyperparameter). This is significantly better than checking all possible task-host combinations, \textit{viz}, $O(pq)$ in the typical case where $p > q$.

\section{Performance Evaluation}
\label{sec:experiments}

We now describe how we evaluate the HUNTER scheduler and compare it against the state-of-the-art baselines: PADQN, CRUZE, MITEC, ANN, SDAE-MMQ and HDIC as described in Section~\ref{sec:related}.

\subsection{Evaluation Setup}
We have tested our proposed approach in both real and simulated cloud environments using the COSCO framework~\cite{tuli2021cosco} and the CloudSim toolkit~\cite{calheiros2011cloudsim}. We keep the size of the scheduling interval as 5 minutes or 300 seconds and run our experiments for 100 scheduling intervals to generate QoS results. For statistical significance, we average over 5 runs. The first is a \textit{physical setup} where we have used 10 Azure VMs in a distributed cloud setup as described below.
\begin{compactitem}
    \item \textbf{Private Cloud.} 6 Azure machines, four of type B2s (dual-core CPU with 4 GB RAM) and two of type B4ms (quad-core with 16 GB RAM). They were instantiated in the London, UK Azure datacenter.
    \item \textbf{Public Cloud.} 4 Azure machines, two of type B4ms (quad-core with 16 GB RAM) and two of type B8ms (octa-core with 32 GB RAM). They were instantiated in the Virginia, USA Azure datacenter.
\end{compactitem}
We also tested on a \textit{simulated} platform with 5 times the instances as described above to give a total of 50 hosts. The former allows more accurate testing of our approach while the latter allows large-scale experimentation. We use the SPEC power benchmarks to determine the energy consumption of the datacenters as done in prior work~\cite{tuli2021cosco}. We used the $R$ and $C$ values in~\eqref{eq:temp} as 0.5 and 0.03 based on prior work~\cite{wolf2016physics}. We followed the same implementation details as in prior work~\cite{tuli2021cosco, basu2019learn, alwasel2020iotsim}.

\subsection{Workloads}
For our physical experiments we use the \textit{DeFog} benchmarking applications for their diverse and non-stationary workloads~\cite{mcchesney2019defog}. DeFog consists of various compute intensive AI applications like \textit{Yolo}, \textit{PocketSphinx} and \textit{Aeneas}. The include workloads in the form of object detection in images, natural language processing, audio-text synchronization and speech recognition. We encapsulate these workloads as Docker containers to execute in our cloud servers. At the start of each scheduling interval we create $Poisson(\lambda)$ jobs with $\lambda = 1.2$. The jobs are sampled uniformly from the three applications of \textit{Yolo}, \textit{PocketSphinx} and \textit{Aeneas}. We divide the input batch of each job into 3 to 5 parts and send them to separate containers (each container acts as a task). 

\begin{figure*}[t]
    \centering \setlength{\belowcaptionskip}{-10pt}
    \subfigure[Energy Consumption]{
    \includegraphics[height=.23\textwidth]{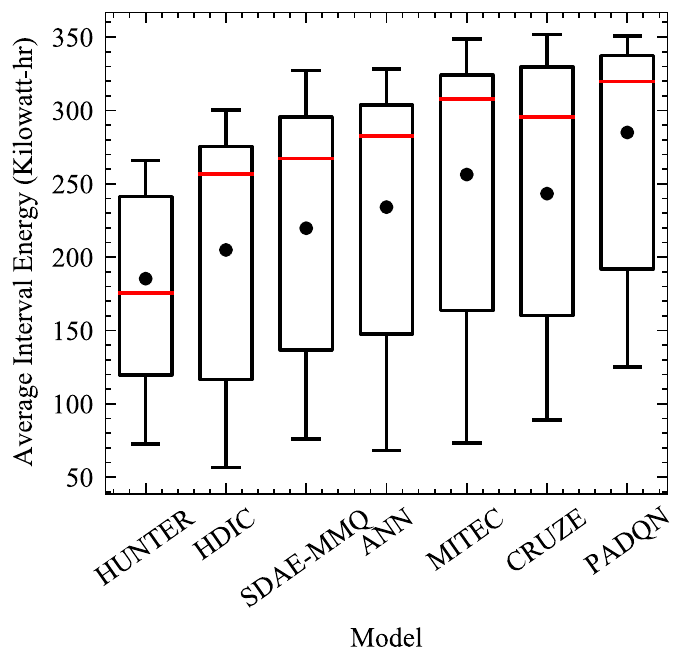}
    \label{fig:f_energy}
    }
    \subfigure[Temperature]{
    \includegraphics[height=.23\textwidth]{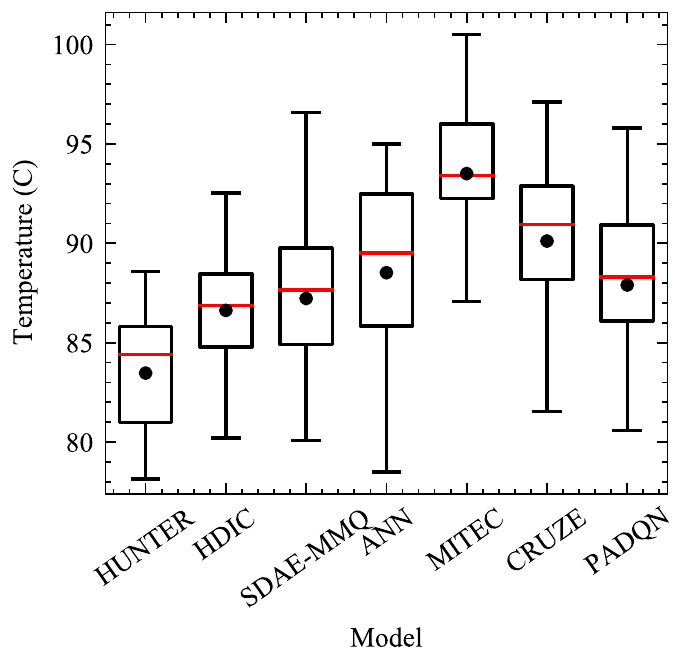}
    \label{fig:f_temp}
    }
    \subfigure[CPU Utilization]{
    \includegraphics[height=.23\textwidth]{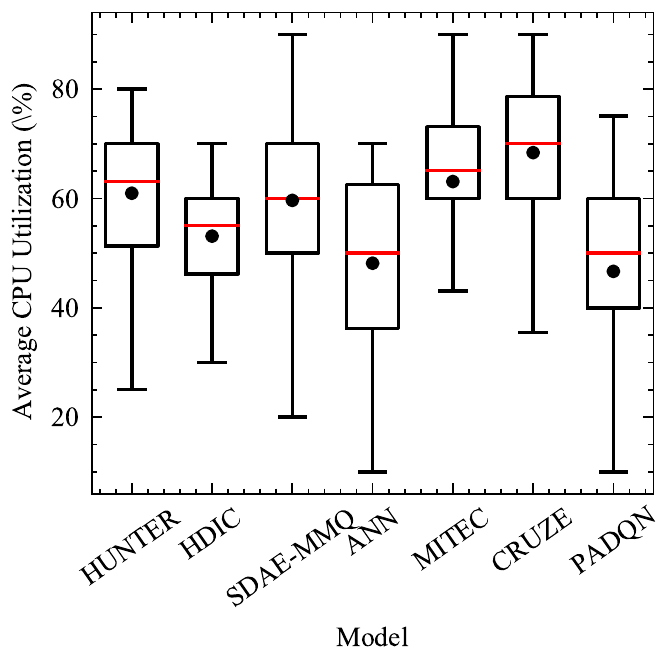}
    \label{fig:f_cpu}
    }
    \subfigure[RAM Utilization]{
    \includegraphics[height=.23\textwidth]{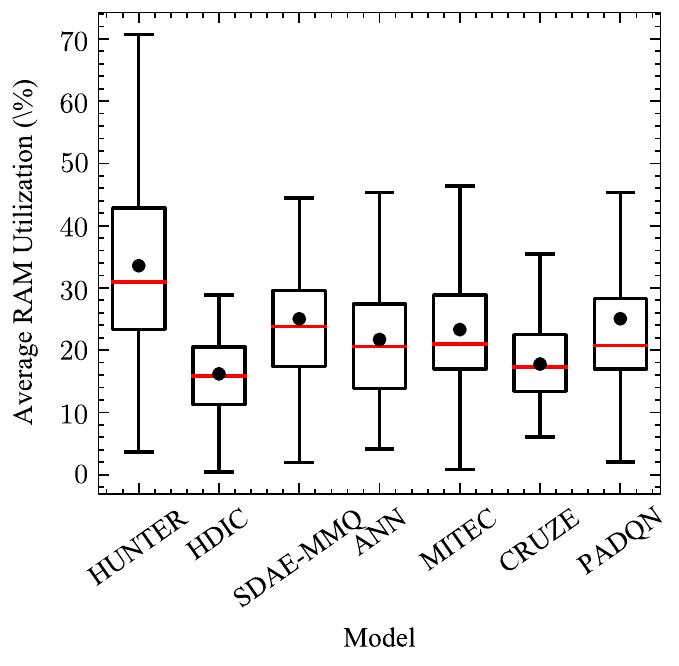}
    \label{fig:f_ram}
    }\\
    \subfigure[SLA Violations]{
    \includegraphics[height=.23\textwidth]{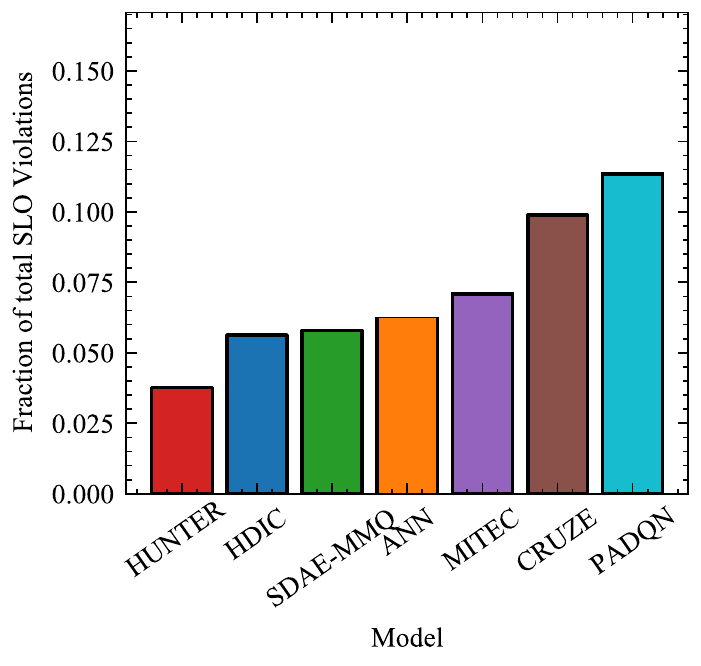}
    \label{fig:f_sla}
    }
    \subfigure[Fairness]{
    \includegraphics[height=.23\textwidth]{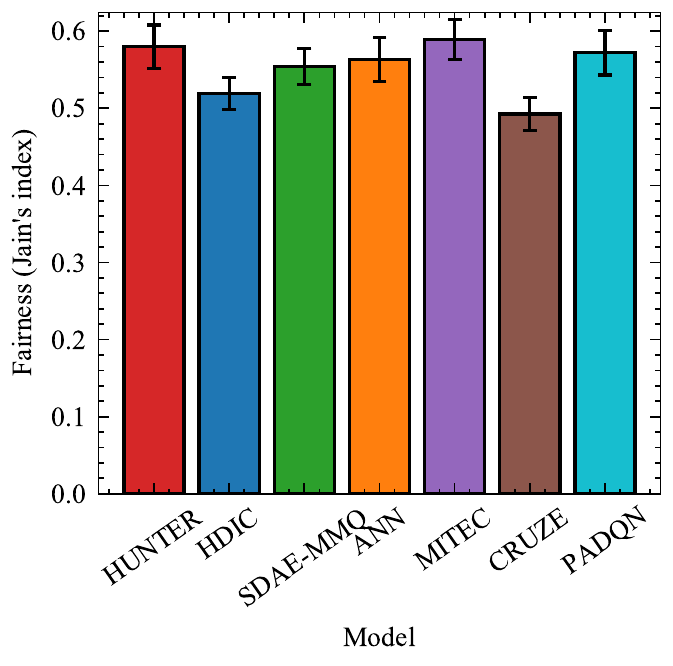}
    \label{fig:f_fairness}
    }
    \subfigure[Completed Tasks]{
    \includegraphics[height=.23\textwidth]{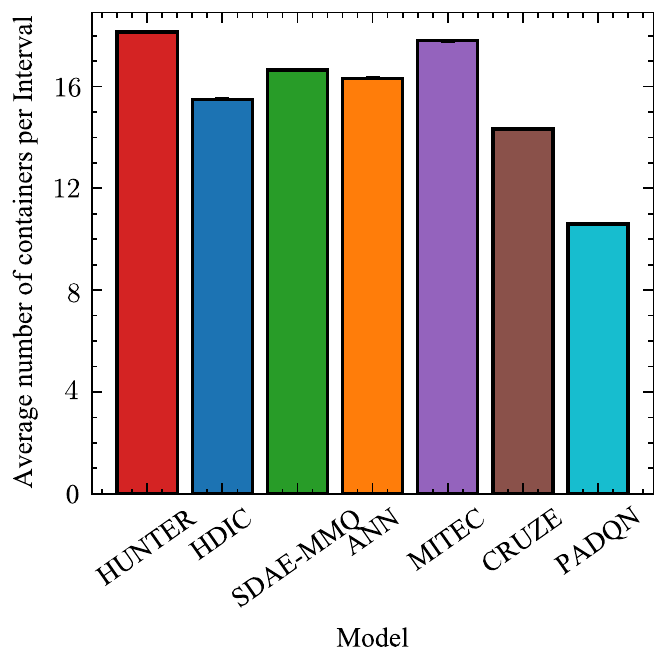}
    \label{fig:f_completed}
    }
    \subfigure[Cost]{
    \includegraphics[height=.23\textwidth]{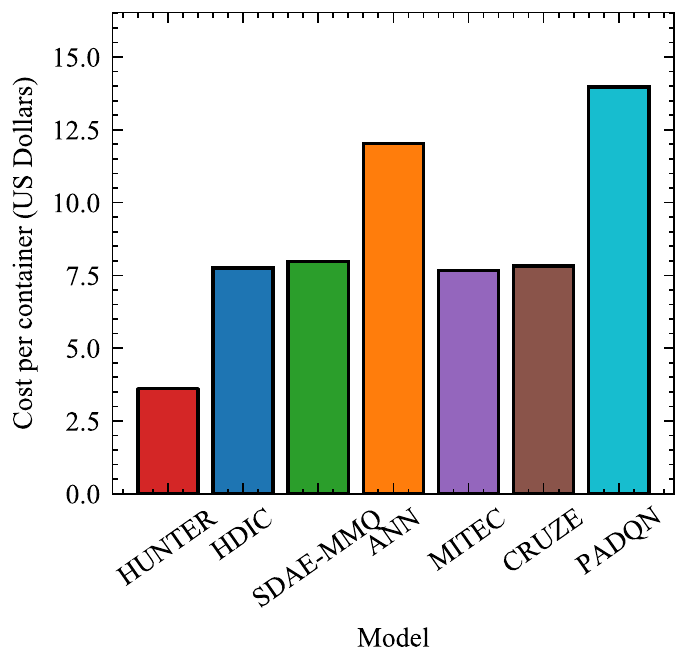}
    \label{fig:f_cost}
    }\\
    \subfigure[Scheduling Time]{
    \includegraphics[height=.215\textwidth]{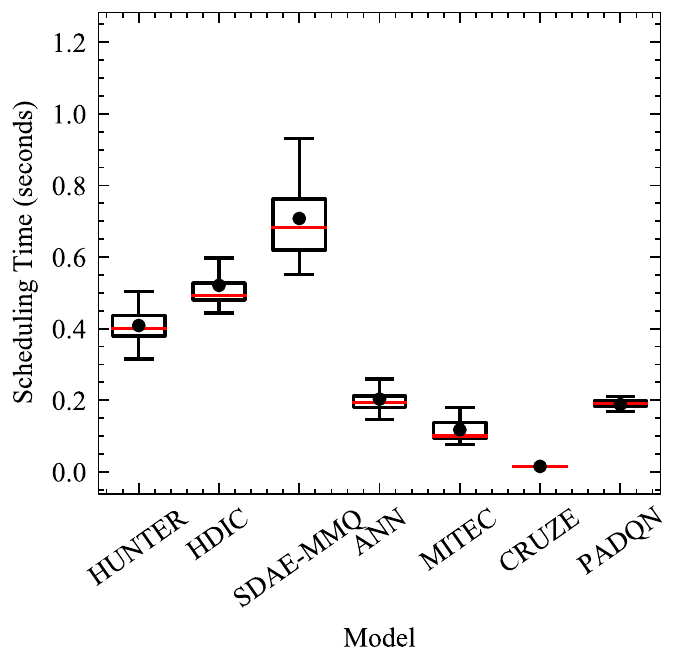}
    \label{fig:f_scheduling_time}
    }
    \subfigure[Wait Time]{
    \includegraphics[height=.215\textwidth]{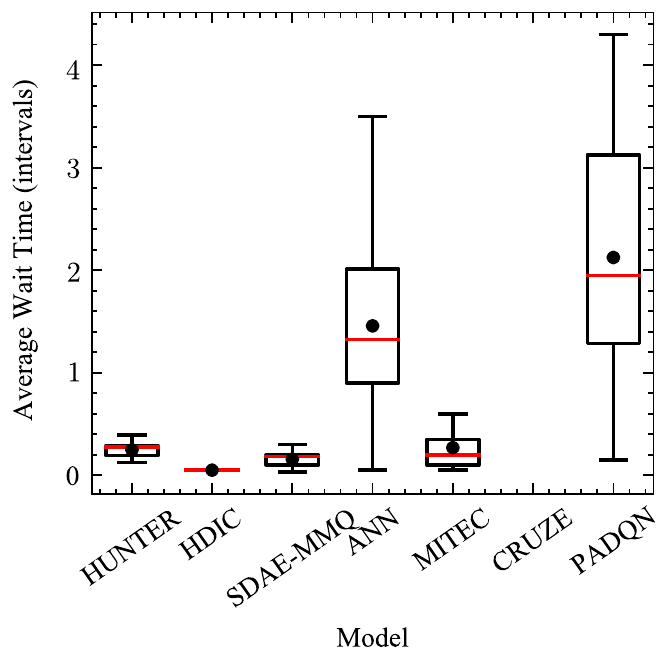}
    \label{fig:f_wait_time}
    }
    \subfigure[Migration Time]{
    \includegraphics[height=.215\textwidth]{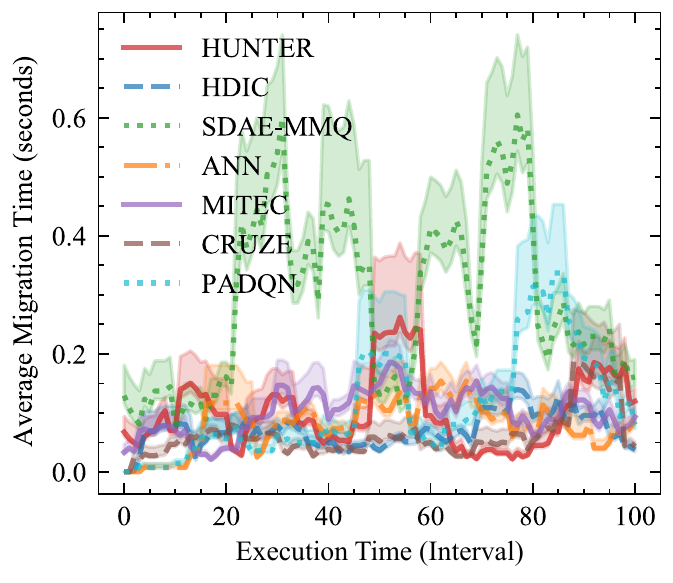}
    \label{fig:f_migration_time}
    }
    \subfigure[Migration Count]{
    \includegraphics[height=.215\textwidth]{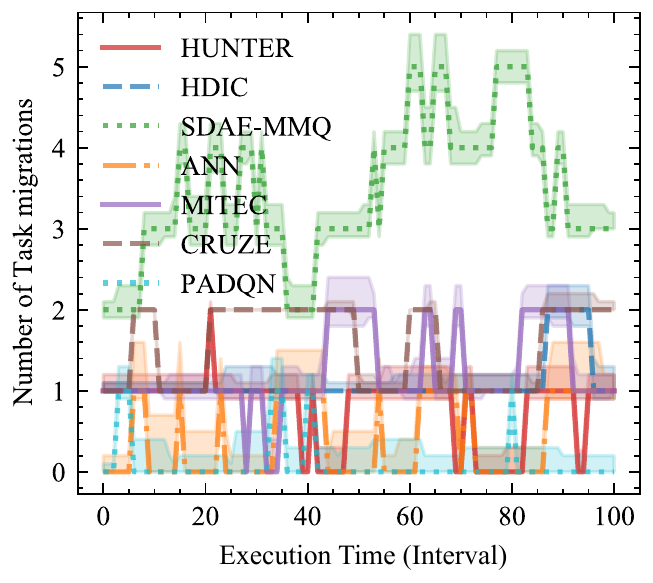}
    \label{fig:f_migrations}
    }
    \caption{Comparison of HUNTER against baselines on physical setup with 10 hosts}
    \label{fig:framework_results}
\end{figure*}

In our simulated setup, we use the popular dynamic traces from the BitBrain dataset to emulate a large-scale execution~\cite{shen2015statistical}. The dataset consists of performance metrics of more than a thousand hosts in a heterogeneous CDC. These traces are collected from the Alibaba distributed datacenter, which is very popular for providing services to perform business computations and manage hosting of industrial applications. Insurers (Aegon), credit card operators (ICS) and many major banks (ING) are the main customers of this datacenter. Further, various financial computing applications (e.g. Algorithmics and Towers Watson) related to credit worthiness domain are hosted here. Moreover, traces are divided into two categories: \textit{Rnd} and \textit{fastStorage}. To allow diverse workloads, we use both traces in our experiments. These traces consist of time-series models of CPU, RAM, Disk and Bandwidth utilization characteristics. As in the physical setup, we create jobs using the the $Poisson(5)$ distribution with each job is sampled uniformly at random from the \textit{Rnd} and \textit{fastStorage} categories and has 3 to 5 tasks. the $\lambda$ parameter is chosen based on prior work~\cite{tuli2021cosco}.

\subsection{Model Training}
\label{sec:training}
The GGCN model takes as an input, the utilization matrix of the active tasks and the capacity matrix of the target hosts. This includes the metrics like CPU, RAM, Disk and Network Bandwidth. We also include the SLA deadline as part of the task utilization matrix. To train the model we first run a random scheduler to cover as much of the state space as possible. We run this for a 1000 scheduling intervals and create a dataset of the form $\{([S, D, T], O)\}$. 

\textbf{Details for Reproducibility:}  We pass the input through a 4 layer GGCN model with 64 nodes each and initialize the hidden state of the GRU by a zero vector. We use AdamW optimizer with a learning rate of $10^{-4}$ to train our model and use early-stopping as our convergence criterion~\cite{adamw}. All model training and experiments were performed on a system with configuration: Intel i7-10700K CPU, 64GB RAM, Nvidia GTX 1060 and Windows 11 OS.

\begin{figure*}[t]
    \centering \setlength{\belowcaptionskip}{-10pt}
    \subfigure[Energy Consumption]{
    \includegraphics[height=.23\textwidth]{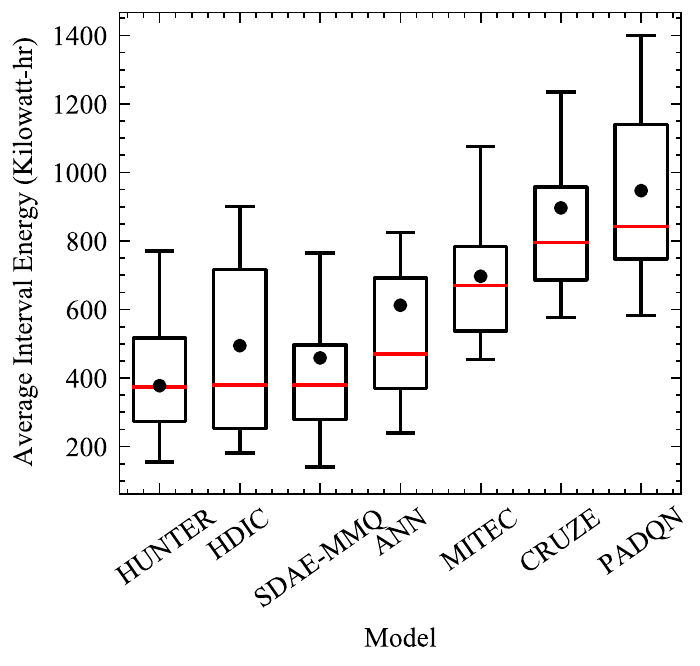}
    \label{fig:s_energy}
    }
    \subfigure[Temperature]{
    \includegraphics[height=.23\textwidth]{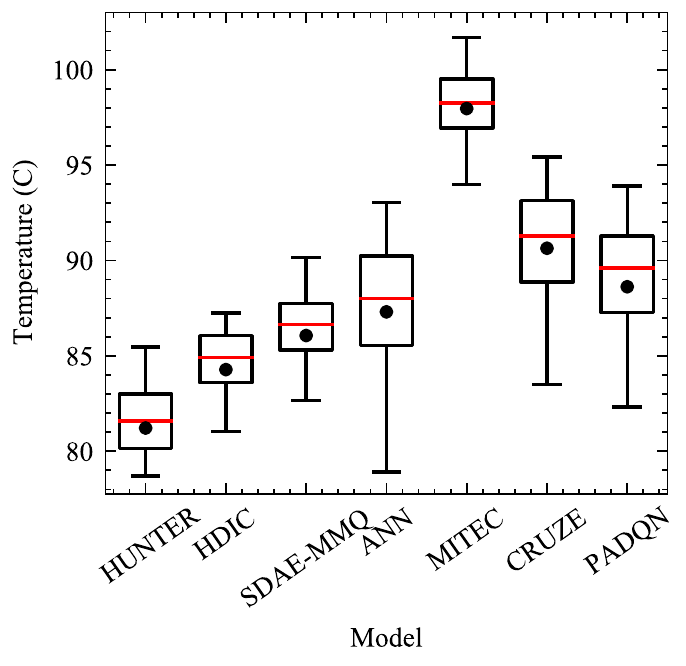}
    \label{fig:s_temp}
    }
    \subfigure[CPU Utilization]{
    \includegraphics[height=.23\textwidth]{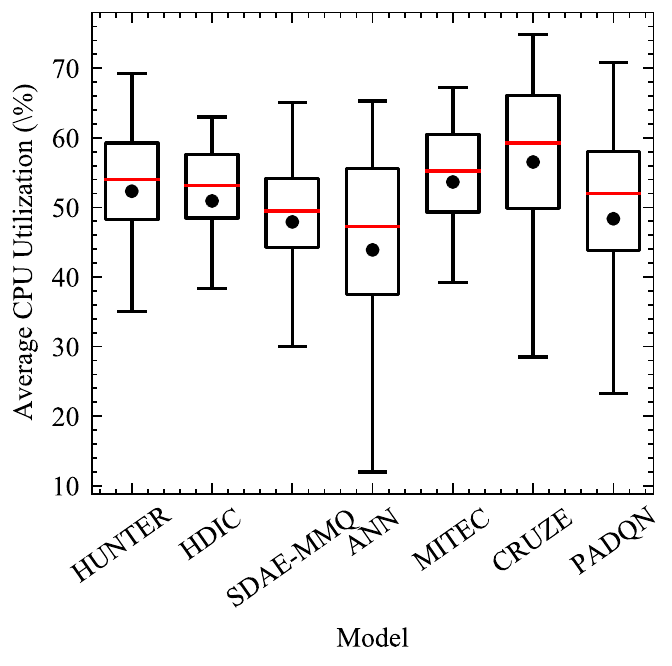}
    \label{fig:s_cpu}
    }
    \subfigure[RAM Utilization]{
    \includegraphics[height=.23\textwidth]{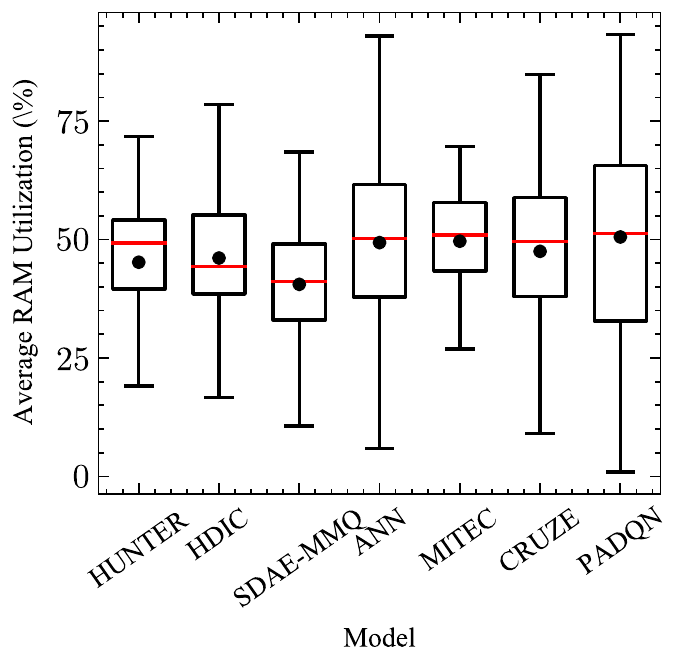}
    \label{fig:s_ram}
    }\\
    \subfigure[SLA Violations]{
    \includegraphics[height=.23\textwidth]{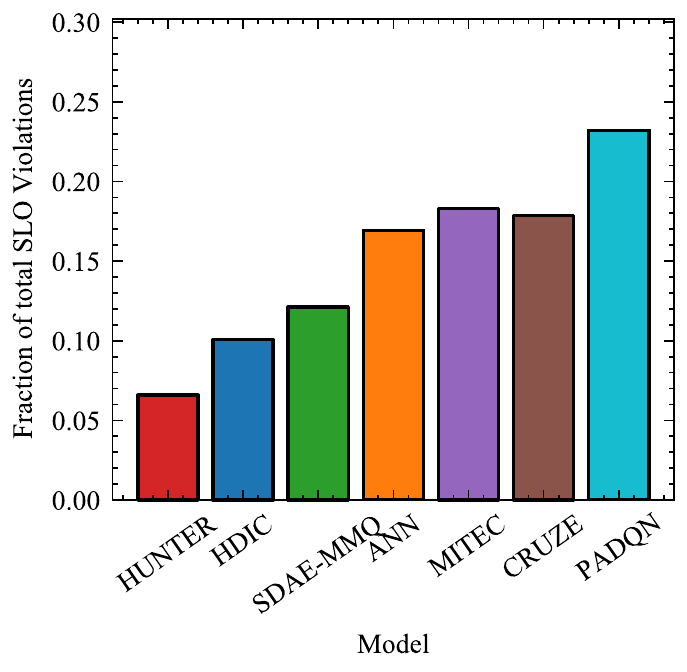}
    \label{fig:s_sla}
    }
    \subfigure[Fairness]{
    \includegraphics[height=.23\textwidth]{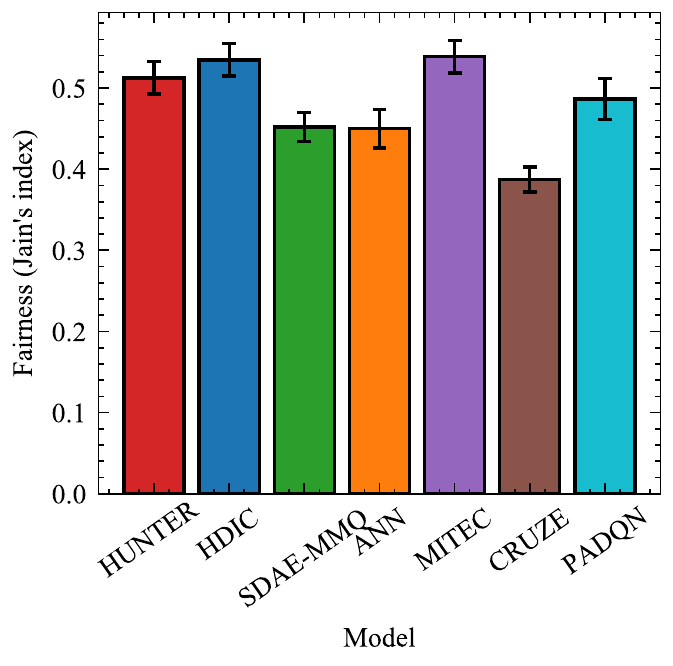}
    \label{fig:s_fairness}
    }
    \subfigure[Completed Tasks]{
    \includegraphics[height=.23\textwidth]{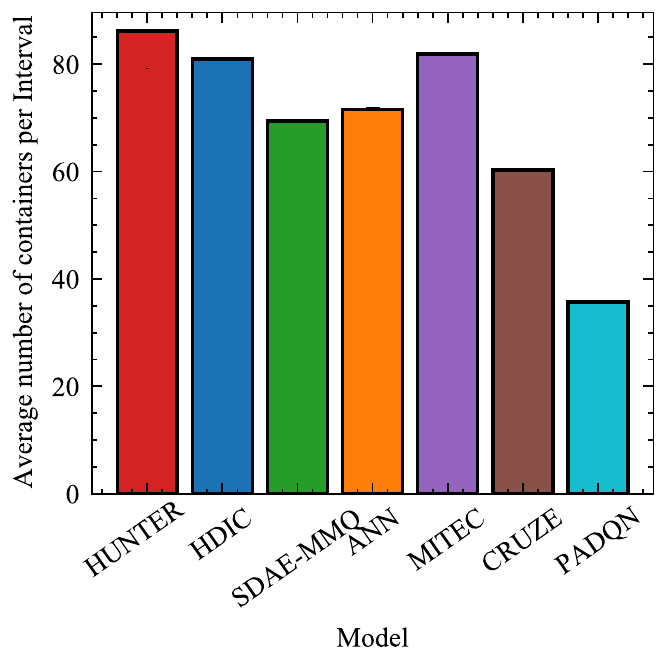}
    \label{fig:s_completed}
    }
    \subfigure[Cost]{
    \includegraphics[height=.23\textwidth]{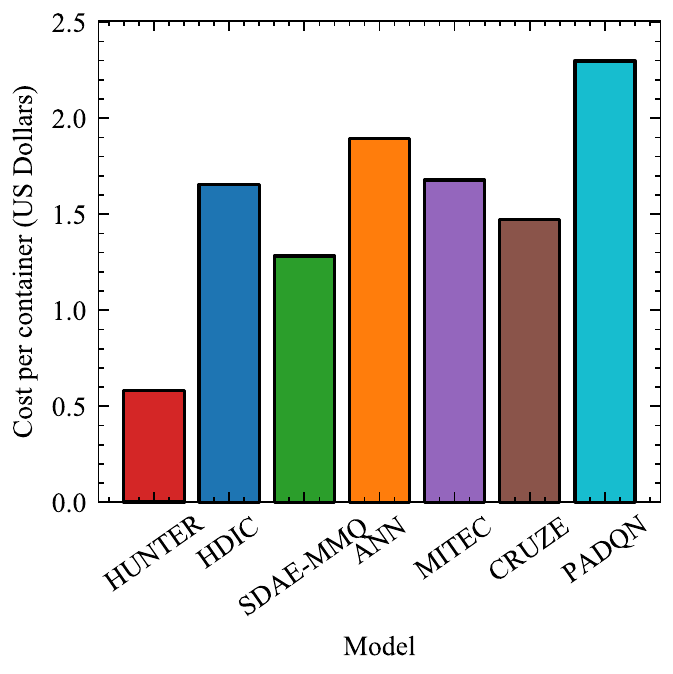}
    \label{fig:s_cost}
    }\\
    \subfigure[Scheduling Time]{
    \includegraphics[height=.215\textwidth]{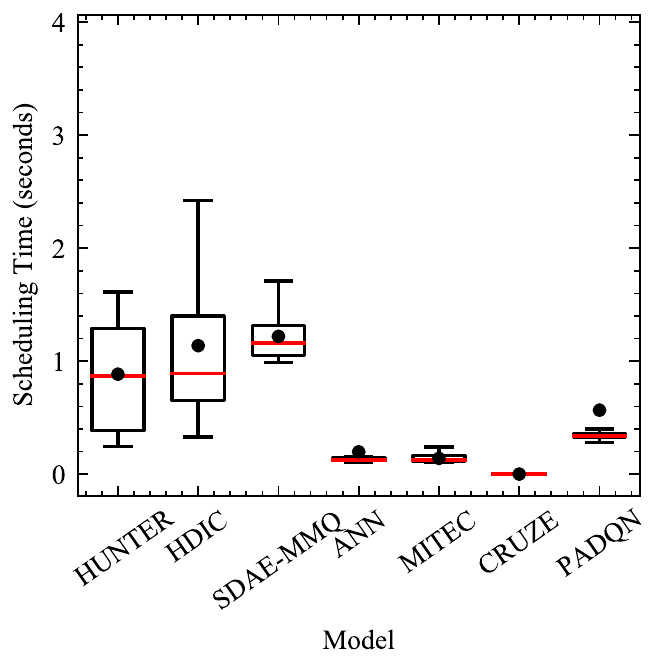}
    \label{fig:s_scheduling_time}
    }
    \subfigure[Wait Time]{
    \includegraphics[height=.215\textwidth]{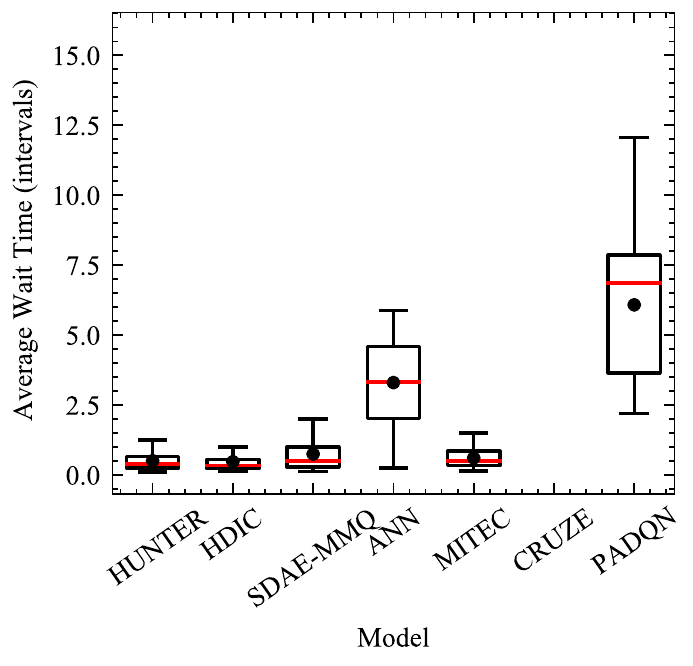}
    \label{fig:s_wait_time}
    }
    \subfigure[Migration Time]{
    \includegraphics[height=.215\textwidth]{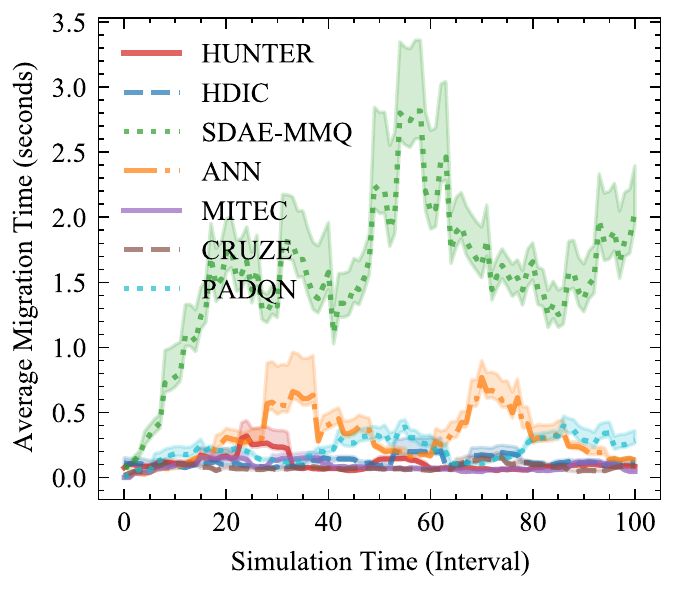}
    \label{fig:s_migration_time}
    }
    \subfigure[Migration Count]{
    \includegraphics[height=.215\textwidth]{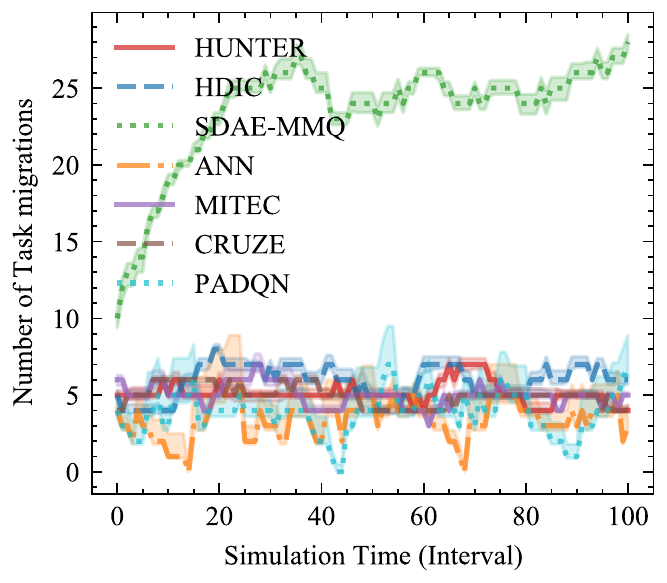}
    \label{fig:s_migrations}
    }
    \caption{Comparison of HUNTER against baselines on simulator with 50 hosts}
    \label{fig:simulator_results}
\end{figure*}

\subsection{Evaluation Metrics}
To compare the proposed HUNTER method against the baseline methods, we use the following metrics:
\begin{compactitem}
    \item \textit{Energy consumption} given as $E_{total}$ in \eqref{eq:energy}.
    \item \textit{SLA Violations} which is given as  \[\frac{\sum_i SLAV_i }{\sum_i j_i},\] where $SLAV_i$ is the 1 if SLA of job $j_i$ is violated else 0. 
    \item \textit{Average Response Time} which is the mean response time for all completed jobs in an experiment
    \item \textit{Datacenter Temperature} given by \eqref{eq:temp}.
    \item {\textit{Cost}} is given by $Time \times Price$. We use the Microsoft Azure pricing calculator to obtain the cost of execution per hour (in US Dollars) \url{https://azure.microsoft.com/en-gb/pricing/calculator/}.
    \item {\textit{Fairness}} is given by the Jain's fairness index~\cite{tuli2021cosco}.
    \item {\textit{Scheduling Time}} is the average time to generate a scheduling decision.
    \item {\textit{Wait Time}} is given as the average time a job spends in the waiting queue.
    \item {\textit{Migration Time}} is the average time a task spends in container migration.
\end{compactitem}

\begin{figure*}[t]
    \centering
    \includegraphics[width=.7\textwidth]{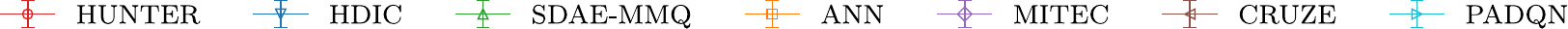}\\
    \subfigure[Energy]{
    \includegraphics[height=0.202\textwidth]{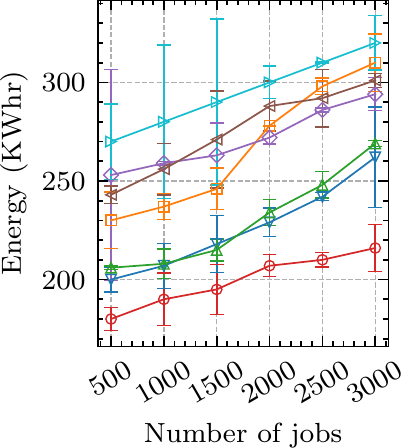}
    \label{fig:sens_f_energy}
    }
    \subfigure[Temperature]{
    \includegraphics[height=0.202\textwidth]{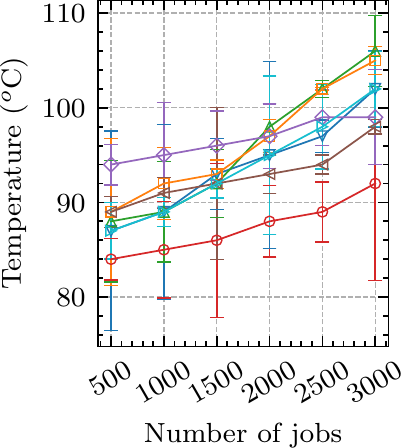}
    \label{fig:sens_f_temp}
    }
    \subfigure[SLA Violations]{
    \includegraphics[height=0.202\textwidth]{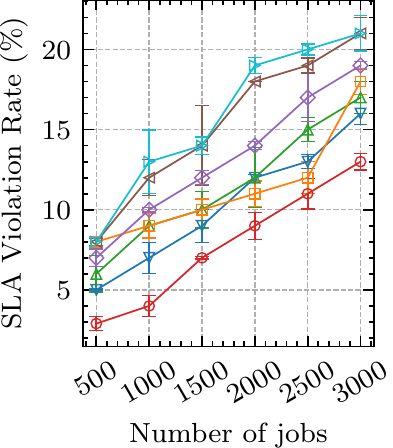}
    \label{fig:sens_f_sla}
    }
    \subfigure[Cost]{
    \includegraphics[height=0.202\textwidth]{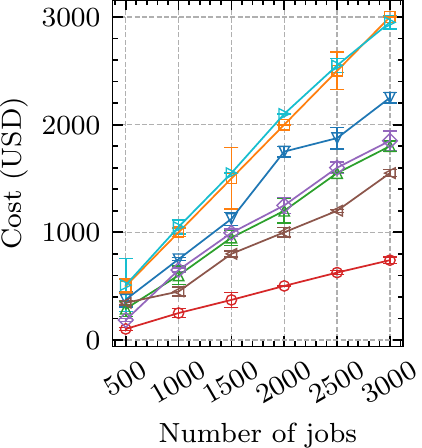}
    \label{fig:sens_f_cost}
    }
    \subfigure[Scheduling Time]{
    \includegraphics[height=0.202\textwidth]{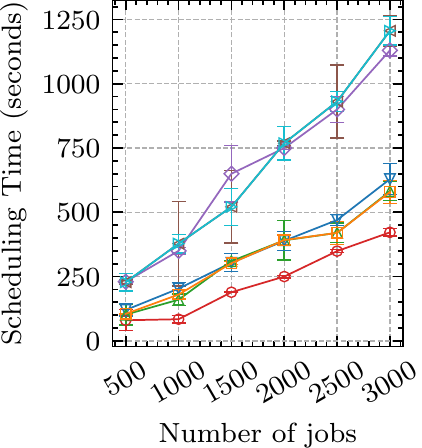}
    \label{fig:sens_f_sched}
    }
    \caption{Sensitivity analysis of HUNTER and baselines with increasing number of workloads on physical setup with 10 hosts.}
    \label{fig:f_sensitivity}
\end{figure*}
\begin{figure*}[t]
    \centering \setlength{\belowcaptionskip}{-10pt}
    \includegraphics[width=.7\textwidth]{images/legend.pdf}\\
    \subfigure[Energy]{
    \includegraphics[height=0.202\textwidth]{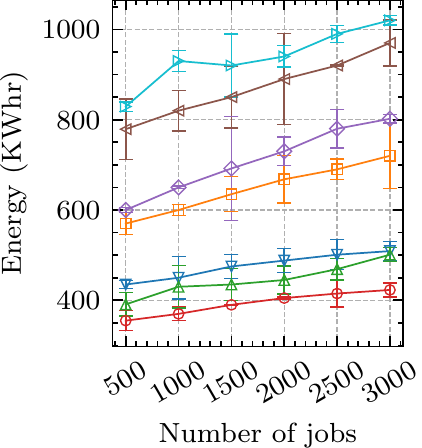}
    \label{fig:sens_s_energy}
    }
    \subfigure[Temperature]{
    \includegraphics[height=0.202\textwidth]{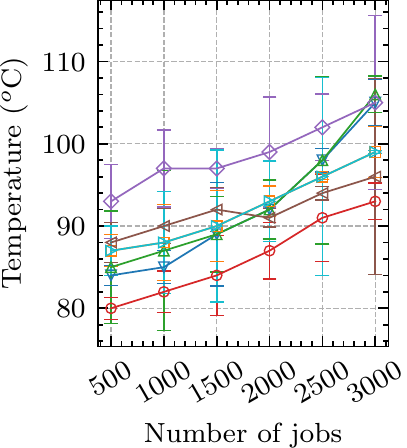}
    \label{fig:sens_s_temp}
    }
    \subfigure[SLA Violations]{
    \includegraphics[height=0.202\textwidth]{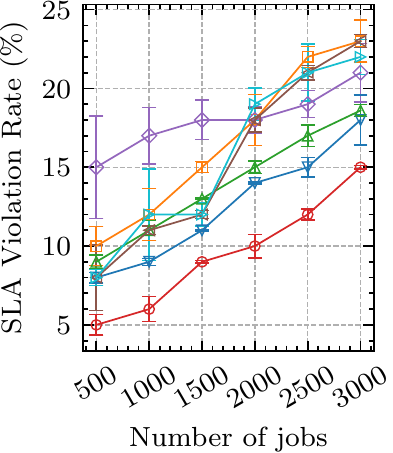}
    \label{fig:sens_s_sla}
    }
    \subfigure[Cost]{
    \includegraphics[height=0.202\textwidth]{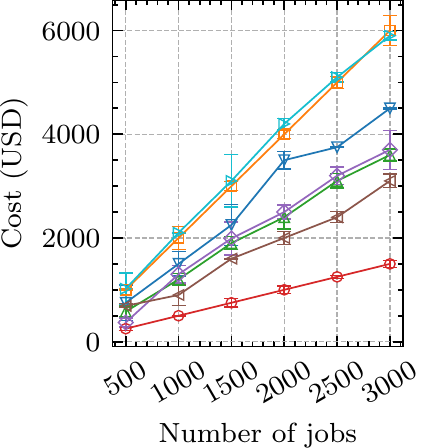}
    \label{fig:sens_s_cost}
    }
    \subfigure[Scheduling Time]{
    \includegraphics[height=0.202\textwidth]{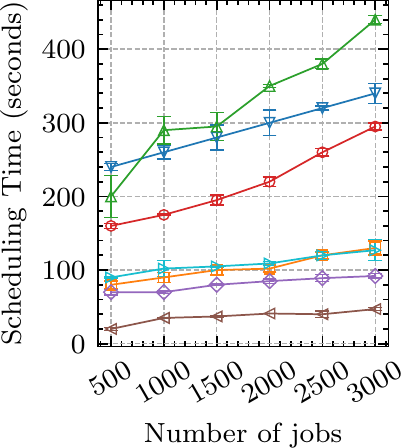}
    \label{fig:sens_s_sched}
    }
    \caption{Sensitivity analysis of HUNTER and baselines with increasing number of workloads on simulated setup with 50 hosts.}
    \label{fig:s_sensitivity}
\end{figure*}
\begin{figure*}[]
    \centering
    \includegraphics[width=.3\textwidth]{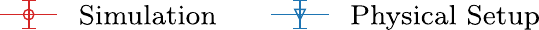}\\
    \subfigure[Energy]{
    \includegraphics[height=0.202\textwidth]{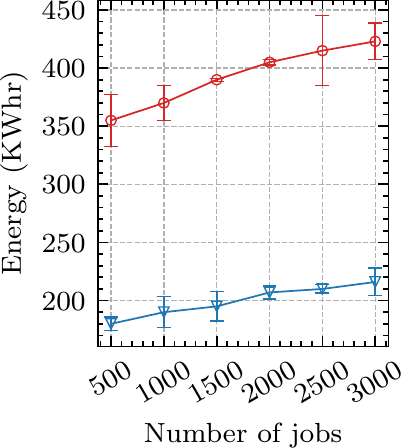}
    \label{fig:comp_energy}
    }
    \subfigure[Temperature]{
    \includegraphics[height=0.202\textwidth]{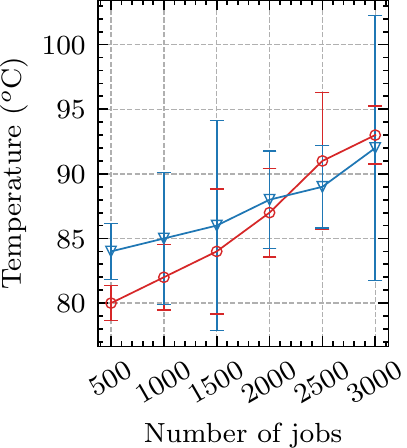}
    \label{fig:comp_temp}
    }
    \subfigure[SLA Violations]{
    \includegraphics[height=0.202\textwidth]{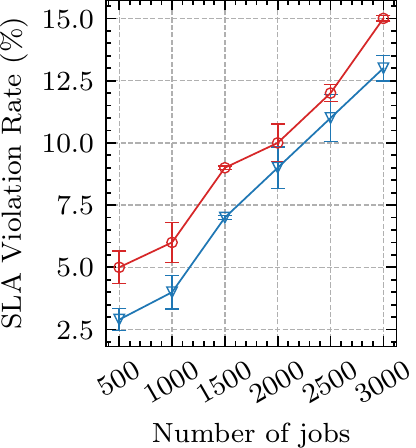}
    \label{fig:comp_sla}
    }
    \subfigure[Cost]{
    \includegraphics[height=0.202\textwidth]{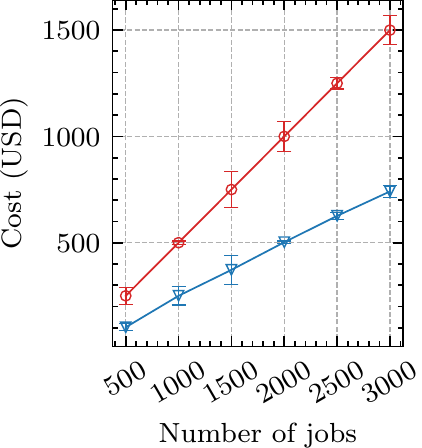}
    \label{fig:comp_cost}
    }
    \subfigure[Scheduling Time]{
    \includegraphics[height=0.202\textwidth]{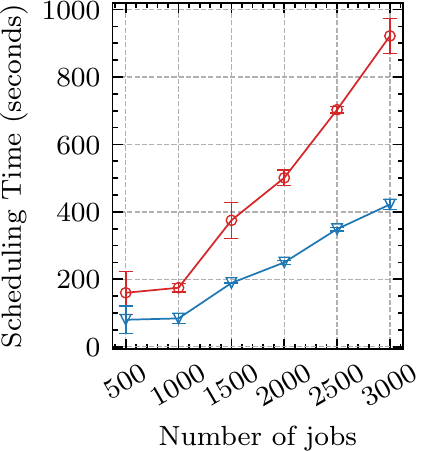}
    \label{fig:comp_sched}
    }
    \caption{Comparison of QoS metrics using the HUNTER scheduler on simulated and physical platforms.}
    \label{fig:comparison}
\end{figure*}

\begin{figure*}[t]
    \centering \setlength{\belowcaptionskip}{-10pt}
    \includegraphics[width=.7\textwidth]{images/legend.pdf}\\
    \subfigure[Energy]{
    \includegraphics[height=0.21\textwidth]{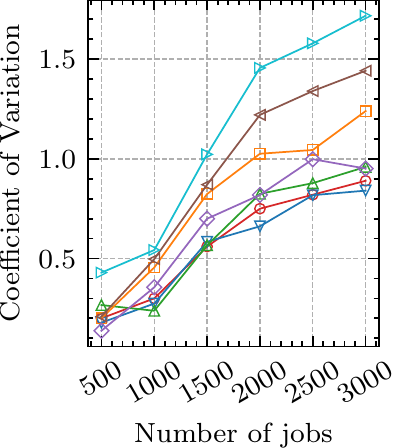}
    \label{fig:cov_energy}
    }
    \subfigure[Temperature]{
    \includegraphics[height=0.21\textwidth]{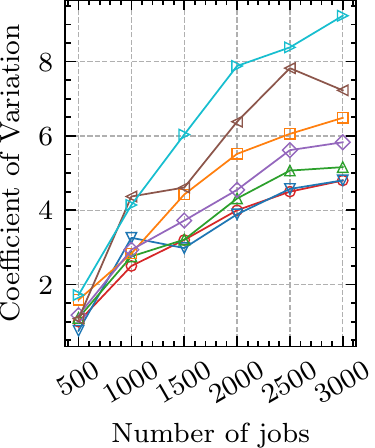}
    \label{fig:cov_temp}
    }
    \subfigure[SLA Violations]{
    \includegraphics[height=0.21\textwidth]{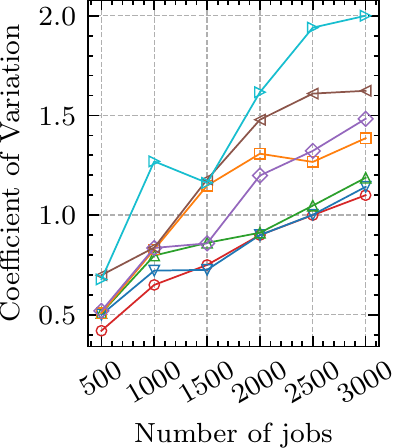}
    \label{fig:cov_sla}
    }
    \subfigure[Cost]{
    \includegraphics[height=0.21\textwidth]{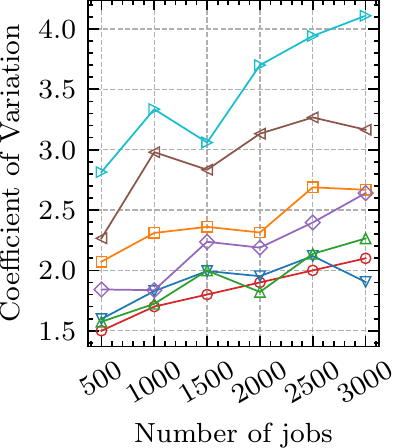}
    \label{fig:cov_cost}
    }
    \subfigure[Scheduling Time]{
    \includegraphics[height=0.21\textwidth]{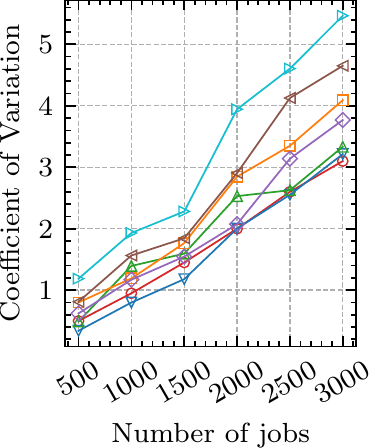}
    \label{fig:cov_sched}
    }
    \caption{Coefficient of Variance for various metrics on the simulated platform.}
    \label{fig:cov}
\end{figure*}

\subsection{QoS Results}
Figures~\ref{fig:framework_results} and~\ref{fig:simulator_results} show the QoS parameters on the COSCO framework and CloudSim simulator respectively. Figures~\ref{fig:f_energy} and~\ref{fig:s_energy} show the energy consumption in a scheduling interval averaged over the number of tasks. Among the baselines, HDIC and SDAE-MMQ provide the most energy efficient policies. As we go from 10 to 50 hosts, the gaps in the energy consumption among schedulers increase, showing how robust the models are in minimizing energy consumption in large-scale setups. Overall, HUNTER gives the lowest energy consumption, reducing by up to 11.90\% compared to the best baselines (HDIC in physical and SDAE-MMQ in simulated testbeds). This is because of the high number of tasks that complete execution in case of HUNTER and the use of the performance-to-power ratio (Figures~\ref{fig:f_completed} and~\ref{fig:s_completed}). Figures~\ref{fig:f_temp} and~\ref{fig:s_temp} show that HUNTER gives the lowest average temperature for both setups, giving a reduction of up to 3.47\% compared to the best baselines, $\sim3^\circ$C (HDIC). This is because of the thermal-aware attention operation in the GGCN based surrogate model that allows HUNTER to emphasize scheduling for hosts that could act as thermal hostpots.  Figures~\ref{fig:f_cpu},~\ref{fig:f_ram},~\ref{fig:s_ram} and~\ref{fig:s_ram} show the CPU and RAM utilization of all models. All models have similar resource utilization metrics, with some cases where the RAM consumption in the HUNTER approach is quite high. Checking the execution traces in case of the HUNTER scheduler shows that this is due to the strict load-balancing rules to prefer keeping the number of containers in hosts to maintain the highest performance to load. Migrations based on such an approach can, in rare cases, lead to slight resource contention at the cost of minimizing energy or temperature. This is primarily when there are sudden spikes in task resource demands, having a cascading effect on other tasks running in the same host. Avoiding such cases is left as part of the future work.  Figures~\ref{fig:a_sla} and~\ref{fig:s_sla} show that the proposed approach is able to reduce SLA violations by up to 35.41\% compared to the best baselines (HDIC). This is primarily due to the accurate QoS objective prediction, allowing the model to minimize the SLA violation rates by checking the QoS estimate for several placement choices. Figures~\ref{fig:f_fairness} and~\ref{fig:s_fairness} show that all models have comparable fairness index values. Figures~\ref{fig:a_cost} and~\ref{fig:s_cost} show the average cost per task for each model. The HUNTER method has the lowest average cost giving up to 53.86\% compared to the best baseline (SDAE-MMQ). PADQN has very high cost due to the excessive migration overheads as shown by Figures~\ref{fig:f_migration_time}, ~\ref{fig:f_migrations},~\ref{fig:s_migration_time},~\ref{fig:s_migrations}. HUNTER like many other baselines has low wait times (Figs.~\ref{fig:f_wait_time} and~\ref{fig:s_wait_time}). Compared to the best baselines (HDIC and SDAE-MMQ), HUNTER has 42.78\% lower scheduling time (Figs.~\ref{fig:f_scheduling_time} and~\ref{fig:s_scheduling_time}).



\section{Analyses}
\label{sec:analyses}

\subsection{Sensitivity Analysis}

We now show how various models scale with the number of workloads. The results in the previous section were time bound, \textit{i.e.}, for 100 scheduling intervals. Now we show how the QoS parameters vary with the number of workloads (see Figs.~\ref{fig:f_sensitivity} and~\ref{fig:s_sensitivity}). Figures~\ref{fig:sens_f_energy} and~\ref{fig:sens_s_energy} show the variation in the consumption of energy with increasing number of jobs. HUNTER consumes up to 19.8\% less as compared to best baseline models (HDIC and SDAE-MMQ). Overall, the rise in energy consumption with number of jobs for HUNTER is not as high as other baseline methods. This is because HUNTER uses the performance to power profiles of cloud hosts to maintain optimal performance while minimizing energy consumption. 
Figures~\ref{fig:sens_f_temp} and~\ref{fig:sens_s_temp} show the change in the temperature with the variation of job quantity. The value of temperature in HUNTER is 5.5\% less than CRUZE because HUNTER uses CRAC-based cooling management~\cite{chaudhry2015thermal} that avoids overloading and underloading of resources and can switch off idle resources automatically.  Figures~\ref{fig:sens_f_sla} and~\ref{fig:sens_s_sla} show the change in the SLA violation rate with the variation of number of jobs. The value of SLA violation rate in HUNTER is up to 42.12\% lower as compared to the HDIC baseline. Figures~\ref{fig:sens_f_cost} and~\ref{fig:sens_s_cost} show the change in the cost with the variation job count. HUNTER gives up to 63.75\% less cost as compared to CRUZE and SDAE-MMQ. This is primarily due to the optimal performance to load management in the HUNTER scheduler. Figures~\ref{fig:sens_f_sched} and~\ref{fig:sens_s_sched} show the distribution of the scheduling time with the job count. There is a sharp increase in PADQN model as DQN scales poorly with time~\cite{tuli2021cosco}. HUNTER has a higher scheduling time compared to the heuristic based baselines: CRUZE and MITEC. However, compared to the best baselines in terms of energy, temperature and cost, \textit{i.e.}, SDAE-MMQ and HDIC, HUNTER has up to 56.12\% lower scheduling times.

\subsection{Comparison between Simulated and Physical Setups}
We now compare the QoS metrics for the HUNTER approach for the simulated (CloudSim) and physical setups (COSCO) (see Figure~\ref{fig:comparison}). Clearly, all QoS parameters increase with the rise of the number of jobs. Figures~\ref{fig:comp_energy}, \ref{fig:comp_temp}, \ref{fig:comp_sla}, \ref{fig:comp_cost} and \ref{fig:comp_sched} show a  performance comparison of simulated and physical setup for energy consumption, host temperature, SLA violation rate, cost and scheduling time. Naturally, we get higher energy consumption, SLA violation rates and scheduling times for the simulated setup as there are five times the number of hosts in that of the physical setup. COSCO allows us to conduct more accurate experiments which give us less noisy fine-tuning as the model adapts to volatile workloads. This describes why the average rise in the temperature with the number of jobs is lower for the physical setup. Moreover, due to the imprecise computation of the resource utilization metrics for the tasks and hosts, the experiments of CloudSim simulator gives results that have high deviation from the ones conducted on the physical platform. Scalability wise we are able to show that HUNTER is able to scale well when the number of workloads or host machines is large.

\begin{table}[t]
    \centering
    \caption{Comparison of training and inference times (in seconds) between HUNTER and baseline methods on simulated setup with 50 hosts.}
    \begin{tabular}{@{}lcc@{}}
    \toprule
        Model & Training Time & Inference Time \tabularnewline \midrule
        HUNTER &  $908\pm12$ & $0.88\pm0.13$ \tabularnewline 
        HDIC &  $1193\pm68$ & $1.12\pm0.28$ \tabularnewline 
        SDAE-MMQ &  $2058\pm102$ & $1.17\pm0.05$ \tabularnewline 
        ANN &  $102\pm4$ & $0.20\pm0.01$ \tabularnewline 
        MITEC &  - & $0.19\pm0.03$ \tabularnewline 
        CRUZE &  - & $0.01\pm0.01$ \tabularnewline
        PADQN &  $340\pm81$ & $0.23\pm0.02$ \tabularnewline \bottomrule
    \end{tabular}
    \label{tab:comparison}
\end{table}

\begin{figure*}[t]
    \centering \setlength{\belowcaptionskip}{-10pt}
    \includegraphics[width=.55\textwidth]{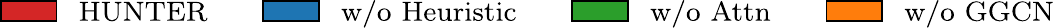}\\
    \subfigure[Prediction Error]{
    \includegraphics[height=.173\textwidth]{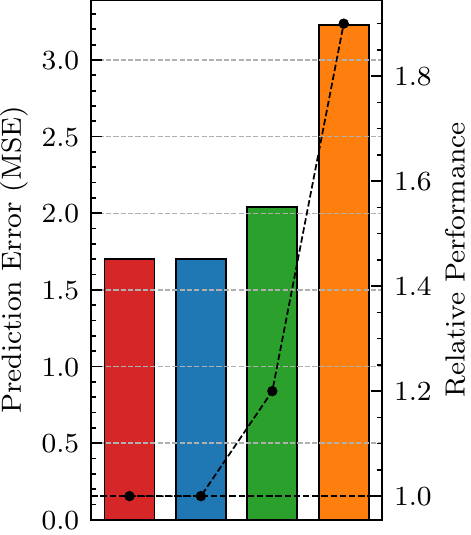}
    \label{fig:a_error}
    }
    \subfigure[Energy]{
    \includegraphics[height=.173\textwidth]{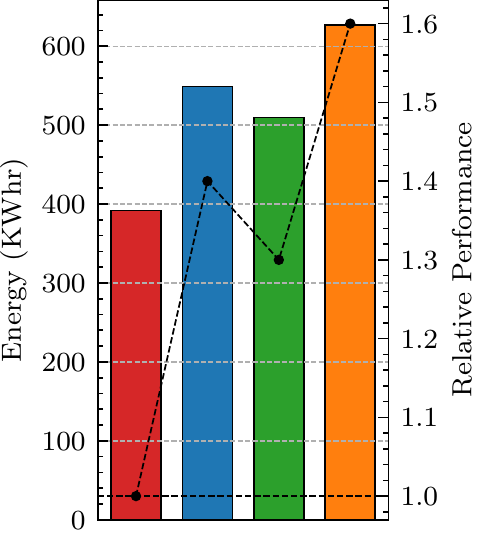}
    \label{fig:a_energy}
    }
    \subfigure[Temperature]{
    \includegraphics[height=.173\textwidth]{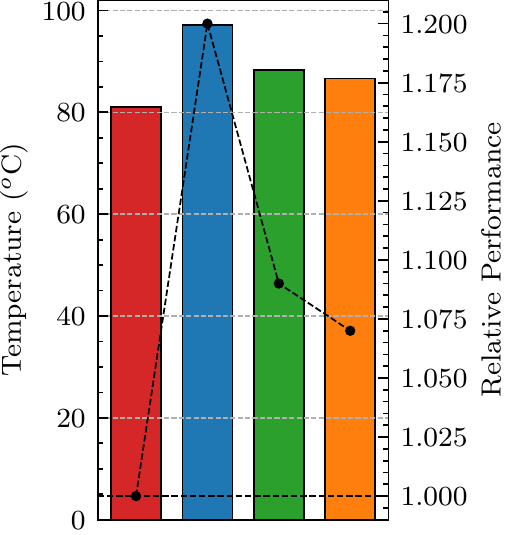}
    \label{fig:a_temp}
    }
    \subfigure[SLA Violations]{
    \includegraphics[height=.173\textwidth]{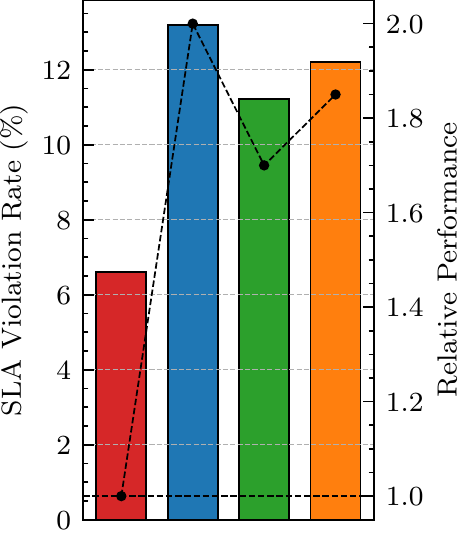}
    \label{fig:a_sla}
    }
    \subfigure[Cost]{
    \includegraphics[height=.173\textwidth]{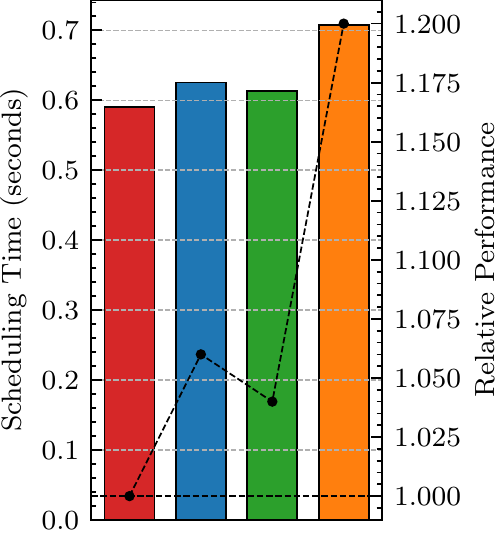}
    \label{fig:a_cost}
    }
    \subfigure[Scheduling Time]{
    \includegraphics[height=.173\textwidth]{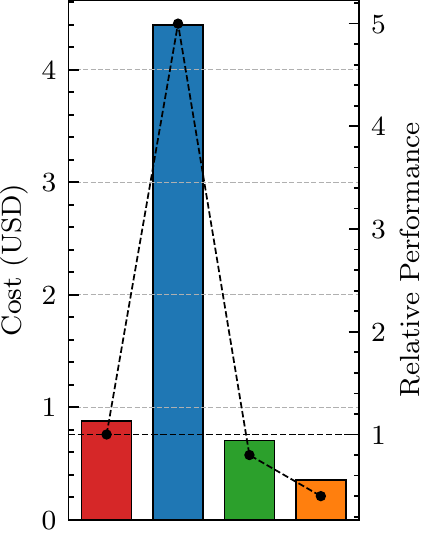}
    \label{fig:a_sched}
    }
    \caption{Ablation Analysis of different model components of HUNTER for the simulated setup with 50 hosts. The bar graphs show absolute values. The line graphs show the performance relative to HUNTER.}
    \label{fig:ablation}
\end{figure*}

\subsection{Analysis of Model Training and Inference Times}
Table~\ref{tab:comparison} compares the training and inference times of the HUNTER approach with the baseline methods. CRUZE and MITEC do not have any training overheads as they run online without training any AI model or neural network. 
We also test the training time for the GGCN model. Compared to various other prior works which rely on deep reinforcement learning (HDIC, SDAE-MMQ, ANN and PADQN) that take up to $2058\pm 102$ seconds, HUNTER takes only $908 \pm 12$ seconds to train its model, giving a training overhead reduction of 55.87\% compared to SDAE-MMQ and 23.88\% compared to HDIC. This is negligible compare to the discrete time interval of $300$ seconds used in our orchestration controllers and hence it is feasible to adopt the HUNTER approach in dynamically changing environments.

When comparing the inference times, the best baselines (HDIC and SDAE-MMQ) have a relatively high inference times of up to 1.17 seconds. HUNTER gives a scheduling time of 0.88 seconds, 24.78\% lower than these baselines.

\subsection{Statistical Analysis}
The Coefficient of Variation (CoV) is used to analyse statistical significance of the experiments as it measures the distribution of the QoS metrics around the mean-value. Moreover, CoV gives an overall analysis of HUNTER's robustness to environment volatility. Figures~\ref{fig:cov_energy}, \ref{fig:cov_temp}, \ref{fig:cov_sla}, \ref{fig:cov_cost} and \ref{fig:cov_sched} show the CoV of energy consumption, temperature, SLA violation rate, execution cost and scheduling time with the increase in number of jobs. The range of CoV is (0.2–0.89\%) for energy consumption, (0.42–1.1\%) for SLA violation rate, (1.5–2.1\%) for cost, (0.5–3.1\%) for scheduling time and (1–6\%) for temperature. HUNTER has comparatively low CoV indicates that the model is able to handle dynamic workloads well and is robust enough to handle environment non-stationarity~\cite{cruze}.

\subsection{Ablation Analysis}
To study the relative importance of each component of the model, we exclude every major one and observe how it affects the performance of the scheduler. An overview of this ablation analysis is given in Figure~\ref{fig:ablation}. First, we consider the HUNTER scheduler without the top-K heuristic and check all task-to-host allocations (\textit{w/o Heuristic} model). Clearly, this gives a much higher scheduling time (Fig.~\ref{fig:a_sched}) and has worse effect on the other QoS metrics due to its high overheads. Second, we consider a model without the thermal-aware attention, \textit{i.e.}, we only use GGCN part of the deep surrogate model (\textit{w/o Attn} model). Here we see that the average temperature increases significantly, also impacting energy and cost. The other model we consider is replacing the GGCN network with a completely feed-forward one (\textit{w/o GGCN} model). Here we see a significant increase in the MSE prediction error (Fig.~\ref{fig:a_error}) leading to higher temperature, cost, SLA violations and energy consumption.

\section{Conclusions and Future Work}
\label{sec:conclusion}

In this paper, we proposed a Gated Graph Convolution Network (GGCN) based holistic resource management scheduling technique called HUNTER. Our scheduler enables energy-efficient utilization of cloud servers and reduces thermal hotspots. HUNTER achieves this by adding cooling specific energy and temperature models, unseen in previous approaches.  Further, using a GGCN based deep surrogate model allows HUNTER to quickly generate QoS estimates, avoiding significant costs in testing various scheduling decisions. HUNTER uses performance to power ratio as a heuristic to effectively balance the load on cloud hosts, giving maximum compute power while reducing energy consumption. This heuristic also allows HUNTER to quickly explore the scheduling search space and quickly converge to a decision.  Extensive experiments on both physical and simulated testbeds show that HUNTER outperforms baselines in most QoS metrics. Furthermore, the small values of the coefficient of variation of energy and temperature indicate that HUNTER is efficient in resource management while handling dynamic workloads. HUNTER optimizes five key performance parameters, \textit{viz}, temperature, energy consumption, cost, SLA violation and time. The experiments demonstrate that the HUNTER performs better than existing AI based (HDIC, SDAE-MMQ, ANN and PADQN) and heuristic algorithm (CRUZE and MITEC) based resource schedulers. 

This work can be extended by factoring in parameters that relate to scalability, security and reliability and their energy ramifications. Future work may also consider how  cooling management can be further enhanced by capturing domain specific tactics for cooling; IoT and Fog/Edge computing reliant domains such as agriculture, healthcare and smart homes are among the candidate application domains to consider. Currently, HUNTER only decides the appropriate placement decisions for tasks; however it can be extended to also decide the AC or fan settings in the cases of deadline constrained or bursty workloads. Finally, HUNTER can use the concept of serverless edge computing to effectively scale applications. 

\section*{Software Availability}
The code is available at \url{https://github.com/imperial-qore/COSCO/tree/ggcn}. The Docker images used in the experiments are available at \url{https://hub.docker.com/u/shreshthtuli}. 

\section*{CRediT authorship contribution statement}
\textbf{Shreshth Tuli:} Conceptualization, Data curation, Investigation, Methodology, Software, Visualization, Validation, Formal analysis, Writing - original draft. \textbf{Sukhpal Singh Gill:} Conceptualization, Data curation, Investigation, Methodology, Software, Visualization, Validation, Formal analysis, Writing - original draft. \textbf{Minxian Xu:} Investigation, Methodology, Writing - original draft.   \textbf{Peter Garraghan:}  Conceptualization, Data curation, Writing - review \& editing.   \textbf{Rami Bahsoon:} Investigation, Methodology, Writing - original draft, Writing - review \& editing.   \textbf{Schahram Dustdar:} Formal analysis and Writing - review \& editing.   \textbf{Rizos Sakellariou:} Methodology and Writing - review \& editing.   \textbf{Omer Rana:} Visualization and Writing - review \& editing.   \textbf{Rajkumar Buyya:} Conceptualization, Writing - original draft and Supervision.   \textbf{Giuliano Casale:} Supervision, Visualization, Writing - review \& editing.   \textbf{Nicholas R. Jennings:} Conceptualization, Writing - original draft, Supervision, Writing - review \& editing.

\section*{Declaration of Competing Interest}
{The authors declare that they have no known competing financial interests or personal relationships that could have appeared to influence the work reported in this paper.}

\section*{Acknowledgments}
{Shreshth Tuli is supported by the President’s Ph.D. Scholarship at the Imperial College London. This research work is partially supported by the EPSRC Research Grant (EP/V007092/1), National Natural Science Foundation of China (62102408), and SIAT Innovation Program for Excellent Young Researchers. The authors thank Muhammad Hassaan Anwar for his suggestions at the initial stages of this work.}


\bibliographystyle{elsarticle-num}
\bibliography{references}
\vfill{}
\begin{wrapfigure}{l}{25mm} 
    \includegraphics[width=1in,height=1.25in,clip,keepaspectratio]{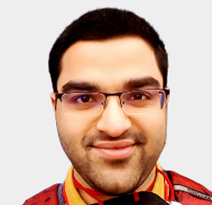}
  \end{wrapfigure}\par
  \textbf{Shreshth Tuli} is an undergraduate student at the Department of Computer Science and Engineering at Indian Institute of Technology - Delhi, India. He is a national level Kishore Vaigyanic Protsahan Yojana (KVPY) scholarship holder for excellence in science and innovation. He is working as a visiting research fellow at the Cloud Computing and Distributed Systems (CLOUDS) Laboratory, Department of Computing and Information Systems, the University of Melbourne, Australia. Most of his projects are focused on developing technologies for future requiring sophisticated hardware-software integration. His research interests include Internet of Things (IoT), Fog Computing, Network Design, Blockchain and deep learning. For further information, visit \url{https://shreshthtuli.github.io/}.\par
\vfill{}
\begin{wrapfigure}{l}{25mm} 
    \includegraphics[width=1in,height=1.25in,clip,keepaspectratio]{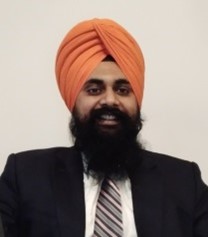}
  \end{wrapfigure}\par
  \textbf{Sukhpal Singh Gill} is a Lecturer (Assistant Professor) in Cloud Computing at School of Electronic Engineering and Computer Science, Queen Mary University of London, UK. Prior to this, Dr. Gill has held positions as a Research Associate at the School of Computing and Communications, Lancaster University, UK and also as a Postdoctoral Research Fellow at CLOUDS Laboratory, The University of Melbourne, Australia. Dr. Gill is serving as an Associate Editor in Wiley ETT and IET Networks Journal. His research interests include Cloud Computing, Fog Computing, Software Engineering, Internet of Things and Healthcare. For further information, please visit \url{http://www.ssgill.me}.\par
\vfill{}
\begin{wrapfigure}{l}{25mm} 
    \includegraphics[width=1in,height=1.25in,clip,keepaspectratio]{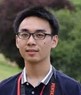}
  \end{wrapfigure}\par
  \textbf{Minxian Xu} is currently an assistant professor at Shenzhen Institutes of Advanced Technology, Chinese Academy of Sciences. He received the BSc degree in 2012 and the MSc degree in 2015, both in software engineering from University of Electronic Science and Technology of China. He obtained his PhD degree from the University of Melbourne in 2019. His research interests include resource scheduling and optimization in cloud computing. He has coauthored 20+ peer-reviewed papers published in prominent international journals and conferences. His Ph.D. Thesis was awarded the 2019 IEEE TCSC Outstanding Ph.D. Dissertation Award. More information can be found at: \url{minxianxu.info}.\par
\vfill{}
\begin{wrapfigure}{l}{25mm} 
    \includegraphics[width=1in,height=1.25in,clip,keepaspectratio]{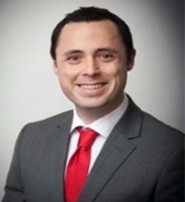}
  \end{wrapfigure}\par
  \textbf{Peter Garraghan} is a Reader and EPSRC Fellow in Distributed Systems. His research expertise is empirically studying and designing high performance, resilient, and sustainable distributed systems at scale (Cloud computing, Machine Learning systems, Datacentres, core network infrastructure) in the face of societal and environmental change. His research places strong emphasis on conducting analysis, design, and evaluation via experimentation on systems both in laboratory and production. Peter has published over 50 articles, has industrial experience building large-scale production distributed systems, and has worked and collaborated internationally with the likes of Alibaba Group, Microsoft, BT, STFC, CONACYT, and the UK datacenter and IoT industry. He is the recipient of the prestigious EPSRC Early-career Fellowship (2021 - 2025), and his research on sustainable computing and future AI systems has featured in the media including the BBC and the Daily Mail.\par
\vfill{}
\begin{wrapfigure}{l}{25mm} 
    \includegraphics[width=1in,height=1.25in,clip,keepaspectratio]{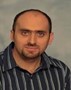}
  \end{wrapfigure}\par
  \textbf{Rami Bahsoon} is a Reader at the School of Computer Science, University of Birmingham, UK. Bahsoon’s research is in the area of software architecture, cloud and services software engineering, self-aware software architectures, self-adaptive and managed software engineering, economics-driven software engineering and technical debt management in software. He co-edited four books on Software Architecture, including Economics-Driven Software Architecture; Software Architecture for Big Data and the Cloud; Aligning Enterprise, System, and Software Architecture. He was a Visiting Scientist at the Software Engineering Institute (SEI), Carnegie Mellon University, USA (June-August 2018) and was the 2018 Melbourne School of Engineering (MSE) Visiting Fellow of The School of Computing and Information Systems, the University of Melbourne (August to Nov 2018). He holds a PhD in Software Engineering from University College London (2006) and was MBA Fellow in Technology at London Business School (2003–2005). He is a fellow of the Royal Society of Arts and Associate Editor of IEEE Software - Software Economies.\par
\vfill{}
\begin{wrapfigure}{l}{25mm} 
    \includegraphics[width=1in,height=1.25in,clip,keepaspectratio]{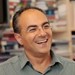}
  \end{wrapfigure}\par
  \textbf{Schahram Dustdar} is Full Professor of Computer Science heading the Research Division of Distributed Systems at the TU Wien, Austria. He is founding co-Editor-in-Chief of the new ACM Transactions on Internet of Things (ACM TIoT) as well as Editor-in-Chief of Computing (Springer). He is an Associate Editor of IEEE TSC IEEE TCC, ACM ToW, and ACM ToIT, as well as on the editorial board of IEEE Internet Computing and IEEE Computer. Dustdar is recipient of the ACM Distinguished Scientist award (2009), the IBM Faculty Award (2012), an elected member of the Academia Europaea: The Academy of Europe, where he is chairman of the Informatics Section, as well as an IEEE Fellow (2016). \par
\vfill{}
\begin{wrapfigure}{l}{25mm} 
    \includegraphics[width=1in,height=1.25in,clip,keepaspectratio]{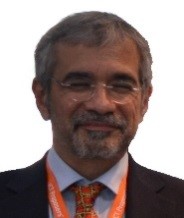}
  \end{wrapfigure}\par
  \textbf{Rizos Sakellariou} obtained his PhD from the University of Manchester in 1997. Since then he held positions with Rice University and the University of Cyprus, while currently he is with the University of Manchester leading a laboratory that carries out research in High-Performance, Parallel and Distributed systems, which over the last 10 years has hosted more than 30 doctoral students, researchers and long-term visitors. Rizos has carried out research on a number of topics related to parallel and distributed computing, with an emphasis on problems stemming from efficient resource utilization and workload allocation and scheduling issues. He has published over 140 research papers, His research has been supported by several UK and EU projects and has been on the Program Committee of over 150 conferences and workshops. \par
\vfill{}
\begin{wrapfigure}{l}{25mm} 
    \includegraphics[width=1in,height=1.25in,clip,keepaspectratio]{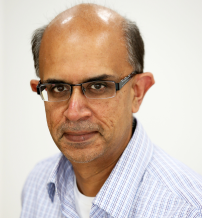}
  \end{wrapfigure}\par
  \textbf{Omer Rana} is a Professor of Performance Engineering in School of Computer Science \& Informatics at Cardiff University and Deputy Director of the Welsh e-Science Centre. He holds a PhD from Imperial College. His research interests extend to three main areas within computer science: problem solving environments, high performance agent systems and novel algorithms for data analysis and management. Moreover, he leads the Complex Systems research group in the School of Computer Science \& Informatics and is director of the ``Internet of Things'' Lab, at Cardiff University. He serves on the Editorial Board of IEEE Transactions on Parallel and Distributed Systems, ACM Transactions on Internet Technology, and ACM Transactions on Autonomous and Adaptive Systems. He has served as a Co-Editor for a number of journals, including Concurrency: Practice and Experience (John Wiley), IEEE Transactions on System, Man, and Cybernetics: Systems, and IEEE Transactions on Cloud Computing. \par
\vfill{}
\begin{wrapfigure}{l}{25mm} 
    \includegraphics[width=1in,height=1.25in,clip,keepaspectratio]{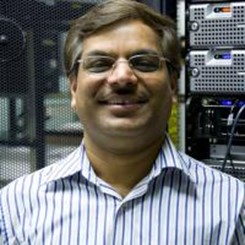}
  \end{wrapfigure}\par
  \textbf{Rajkumar Buyya} is a Redmond Barry Distinguished Professor and Director of the Cloud Computing and Distributed Systems (CLOUDS) Laboratory at the University of Melbourne, Australia. He is also serving as the founding CEO of Manjrasoft, a spin-off company of the University, commercializing its innovations in Cloud Computing. He has authored over 850 publications and seven text books including "Mastering Cloud Computing" published by McGraw Hill, China Machine Press, and Morgan Kaufmann for Indian, Chinese and international markets respectively. He is one of the highly cited authors in computer science and software engineering worldwide (h-index=149, g-index=322, 116,500+ citations).  He served as the founding Editor-in-Chief of the IEEE Transactions on Cloud Computing. He is currently serving as Co-Editor-in-Chief of Journal of Software: Practice and Experience, which was established 50+ years ago. For further information, please visit his cyberhome: \url{www.buyya.com}. \par
\vfill{}
\begin{wrapfigure}{l}{25mm} 
    \includegraphics[width=1in,height=1.25in,clip,keepaspectratio]{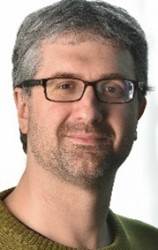}
  \end{wrapfigure}\par
  \textbf{Giuliano Casale} joined the Department of Computing at Imperial College London in 2010, where he is currently a Reader. Previously, he worked as a research scientist and consultant in the capacity planning industry. He teaches and does research in performance engineering and cloud computing, topics on which he has published more than 100 refereed papers. He has served on the technical program committee of over 80 conferences and workshops and as co-chair for several conferences in the area of performance and reliability engineering, such as ACM SIGMETRICS/Performance and IEEE/IFIP DSN. His research work has received multiple awards, recently the best paper award at ACM SIGMETRICS. He serves on the editorial boards of IEEE TNSM and ACM TOMPECS and as current chair of ACM SIGMETRICS.\par
\vfill{}
\begin{wrapfigure}{l}{25mm} 
    \includegraphics[width=1in,height=1.25in,clip,keepaspectratio]{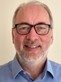}
  \end{wrapfigure}\par
  \textbf{Nicholas R. Jennings} is the Vice-Chancellor and President of Loughborough University. He is an internationally-recognised authority in the areas of AI, autonomous systems, cyber-security and agent-based computing. He is a member of the UK government’s AI Council, the governing body of the Engineering and Physical Sciences Research Council, and chair of the Royal Academy of Engineering’s Policy Committee.  Before Loughborough, he was the Vice-Provost for Research and Enterprise and Professor of Artificial Intelligence at Imperial College London, UK's first Regius Professor of Computer Science (a post bestowed by the monarch to recognise exceptionally high quality research) and the UK Government’s first Chief Scientific Advisor for National Security.\par
\vfill{}

\end{document}